\documentclass[11pt]{book}
\usepackage{array}

\usepackage{amssymb,graphicx,booktabs}

\usepackage[english] {babel}
\usepackage[latin1]{inputenc}
\usepackage[T1]{fontenc}

\usepackage{amsmath, amsthm}
\usepackage{amsfonts,amssymb}

\usepackage[margin=3cm]{geometry}
\usepackage{url}
\usepackage{graphicx}

\usepackage{amssymb}

\newtheorem{result}{Proposition}

\usepackage{latexsym}

\usepackage{bbm}
\newcommand{\ind}{{\mathbbm 1}}

\begin{document}
\renewcommand{\baselinestretch}{1.1}
%\lhead[\fancyplain{} \leftmark]{}
%\chead[]{}
%\rhead[]{\fancyplain{}\rightmark}
%\cfoot{}
%\headrulewidth=0pt
\markright{
%\hbox{\footnotesize\rm Statistica Sinica
%{\footnotesize\bf ??}(200?), 000-000}\hfill
}
\markboth{\hfill{\footnotesize\rm CARDOT, H.  GOGA, C. AND SHEHZAD, M.-A. 
}\hfill}
{\hfill {\footnotesize\rm Calibration on Principal Components} \hfill}
\renewcommand{\thefootnote}{}
$\ $\par
\fontsize{10.95}{14pt plus.8pt minus .6pt}\selectfont
\vspace{0.8pc}
\centerline{\large\bf Calibration and Partial Calibration on Principal Components }
\vspace{2pt}
\centerline{\large\bf when the Number of Auxiliary Variables is Large}
\vspace{.4cm}
\centerline{Cardot, H.$^{(1)}$, Goga, C.$^{(1)}$ and Shehzad, M.-A.$^{(1),(2)}$\footnote{M.-A. Shehzad was supported by the HEC (Higher Education Commission) of Pakistan.}}
\vspace{.4cm}
\centerline{\it $^{(1)}$ IMB, Universit\'e de Bourgogne, DIJON, FRANCE}
 \centerline{\it $^{(2)}$ Bahauddin Zakariya University, MULTAN, PAKISTAN}
\vspace{.55cm}
\fontsize{9}{11.5pt plus.8pt minus .6pt}\selectfont

\begin{quotation}
\noindent {\it Abstract:}
In survey sampling, calibration is a very popular tool used to make total estimators consistent with known totals of auxiliary variables and to reduce variance.
When the number of auxiliary variables is large, calibration on all the variables may lead to estimators of totals whose mean squared error (MSE) is larger than the MSE of the Horvitz-Thompson estimator even if this simple estimator does not take account of the available auxiliary information. We study in this paper a new technique based on dimension reduction through principal components that can be useful in this large dimension context. Calibration is performed on the first principal components, which can be viewed as the synthetic variables containing the most important part of the variability of the auxiliary variables. When some auxiliary variables play  a more important role than the others, the method can be adapted to provide an exact calibration on these important variables. Some asymptotic properties are given in which the number of variables is allowed to tend to infinity with the population size.
A data driven selection criterion of  the number of principal components ensuring that all the sampling weights remain positive is discussed. The methodology of the paper is illustrated, in a multipurpose context, by an application to the estimation of electricity consumption for each day of a  week with the help of 336 auxiliary variables consisting of the past consumption measured every half an hour over the previous week. 
\par

\vspace{9pt}
\par
\end{quotation}\par

\fontsize{10.95}{14pt plus.8pt minus .6pt}\selectfont
\setcounter{chapter}{1}
\setcounter{equation}{0} %-1
\noindent {\bf 1. Introduction}

Since the seminal work by Deville and S\"arndal (1992), calibration is one of the most popular and useful tools to improve Horvitz-Thompson estimators of totals in a 
design-based survey sampling framework. 
Roughly speaking, it consists in looking for a modification of the sampling weights so that the totals, in the population, of the auxiliary variables are perfectly estimated. Performing calibration often leads to total estimators with smaller variances and this technique is routinely  used by several national statistical agencies (see S\"arndal, 2007 for a review).

With the spread of  automatic process for data collection as well as increasing  storage capacities, it is not unusual anymore to have to analyze data coming from very large surveys with many auxiliary variables. In this finite population context, calibration on all the auxiliary variables may lead to estimators whose performances are worse than the simple Horvitz-Thompson estimator even if this latter estimator does not account for any auxiliary information (see \textit{e.g.} Silva and Skinner, 1997). 
Calibration over a very large number of auxiliary variables has been called \textit{over-calibration} by Guggemos and Till\'e (2010). 
Several difficulties arise in this context such as instability of the calibration weights or variance inflation. 
There are different ways of dealing with these issues. One possibility is to choose only a  subset of the auxiliary variables and to consider only the auxiliary variables that are expected to be the more pertinent, avoiding the problem of multicollinearity (see \textit{e.g.} Silva and Skinner, 1997; Chambers, Skinner and Wang, 1999 and Clark and Chambers, 2008).  Another way is to relax the calibration constraints, meaning that the too restrictive requirement of being  exactly calibrated is dropped off and replaced by the requirement of being only approximately calibrated. A class of penalized estimators has been suggested by Bardsley and Chambers (1984) in a model-based setting and extended later by Chambers (1996), Rao and Singh (1997, 2009) and Th\'eberge (2000) in a design-based (or model-assisted) setting. Usually, some auxiliary variables play a role that is more important than the others and it is required that their totals  be estimated exactly. To fulfill this requirement, Bardsley and Chambers (1984) and Guggemos and Till\'e (2010) suggested two different penalized optimization problems which lead in fact to the same system of weights, and by consequence to the same partially penalized estimator (see Goga and Shehzad, 2014).   

We present in this paper another way of dealing with this issue. Our estimator is based on dimension reduction of the auxiliary variables via principal components calibration.  In multivariate statistics, principal component analysis (PCA) is a very  popular tool for reducing the dimension of a set of quantitative variables (see {\it e.g.} Jolliffe, 2002) by transforming the initial data set  into a new set of a few uncorrelated synthetic variables, called principal components (PC), which are linear combinations of the initial variables with the largest variance. The method we suggest consists in reducing the number of auxiliary variables by considering a small number of PC's and by performing calibration on these new synthetic variables. As far as we know, its application in a survey sampling setting is new.

Adopting a model-assisted point of view (see S\"arndal \textit{et. al} 1992), the PCA Calibration approach  can also be viewed as a GREG estimator based on Principal Components Regression (PCR).  
PCR can be very useful to reduce the number of covariates in a linear regression model especially when the regressors are highly correlated (see \textit{e.g} Frank and Friedman,1993). As explained in Jolliffe (2002), even if PCR is a biased estimation method for estimating a regression coefficient, it is useful to overcome the problem of multicollinearity among the regressors.  The method is easy to put into practice with classical softwares used for performing calibration, such as CALMAR used at the French National Statistical Institut (Insee) since, with centered data, these new calibration variables are also centered and their totals are known and equal to zero.

Note that  a natural alternative to principal components regression is partial least squares (PLS) which is also a popular dimension reduction regression technique that can be useful when there is a large number of auxiliary variables that are highly correlated (see for example Swold \textit{et al.} 2001 for a pedagogical presentation of PLS). 
As noted in Frank and Friedman (1993), it often has similar prediction errors to principal components regression. Other model selection techniques such as the Lasso (Tibshirani, 1996) or the elastic net (Zou and Hastie, 2005) can be employed to deal with survey data with large number of auxiliary variables. The main drawback of all  these  model selection techniques is that they would give survey weights that would depend explicitly on the outcome variable. 
Having weights that depend on the outcome variable is generally not desired in surveys, particularly in multipurpose surveys, in which there can be many outcome variables under study. 
Consequently, all these alternative regression techniques that would have given weights depending explicitly on the outcome variable have not been considered in this work.

The paper is structured as follows: we briefly  recall in Section 2 the calibration method as  it was suggested by Deville and S\"arndal (1992) and the problems which may arise when the number of auxiliary variables is large. We introduce in Section 3 the suggested method and we give a model-assisted interpretation. As expected,  with the chi-squared distance, the calibration estimator on PC's  may be written as a GREG-type estimator with respect to the initial auxiliary variables but with a modified regression coefficient estimator. 
When the auxiliary information is not complete, {\it i.e.} the values of the auxiliary variables are only known in the sample, we first estimate the PC's and then we perform calibration on the first estimated principal components (Section 4). In Section 5, under mild assumptions on the sampling design and on the study and auxiliary variables, we prove that the calibration estimator on true PC's as well as on estimated PC's are asymptotical unbiased and consistent. We show also that these estimators are asymptotically equivalent to the generalized difference estimator computed with a PCR estimator. Variance estimators are also presented. We present, in Section~6, how the method can be adapted to provide an exact calibration on variables considered by the survey statistician more important than the other variables. 
Our method is illustrated in Section 7, in a multipurpose and large dimension context, on the estimation of the total electricity consumption for each day of a week with the help of the past consumption measured every half an hour over the previous week. Borrowing ideas from Bardsley and Chambers (1984), a data-driven choice of the tuning parameter is suggested which consists in selecting the largest dimension such that all the estimated principal component weights remain positive.  
Finally, a brief Section 8 gives some  concluding remarks as well as some directions that would deserve further investigation.  The proofs are gathered in an Appendix.

\par

\vspace{1cm}
\setcounter{chapter}{2}
\setcounter{equation}{0} %-1
\noindent {\bf 2. Calibration over a large number of auxiliary variables}\label{chapter2}

We consider the finite population  $U=\{1, \ldots, k, \ldots, N\}$ and we wish to estimate the total $t_y=\sum_{k\in U} y_k,$ 
where $y_k$ is the value of the variable of interest  $\mathcal Y$ for the $k$th unit. Let $s$ be a random sample, with fixed size $n$, drawn from $U$ according to a sampling design that assigns to unit $k$ a known inclusion probability $\pi_k=\rm{Pr}(k\in s)$ satisfying $\pi_k>0$. The corresponding sampling design weight is denoted by  $d_k=1/\pi_k.$ We suppose that $y_k$ is known for all $k\in s$ (complete response). 

 Without auxiliary information, the total $t_y$ is estimated unbiasedly by the Horvitz-Thompson~(HT) estimator $\hat t_{yd}=\sum_{k\in s}d_ky_k.$
Consider now $p$ auxiliary variables, $\mathcal X_1, \ldots, \mathcal X_p$,  and let $\mathbf{x}^{T}_k=(x_{k1}, \ldots, x_{kp})$ be the transposed vector whose elements are the values of the auxiliary variables for the $k$th unit.  The calibration method has been developed by Deville and S\"arndal (1992)  to use as effectively as possible the known population totals of $\mathcal X_j, j=1, \ldots, p$ at the estimation stage. The calibration estimator of $t_y$ is  a weighted  estimator 
\begin{align}
\hat t_{yw} &=\sum_{k\in s}w_k y_k,\label{def:calibest}
\end{align}
whose (calibration) weights  $w_k$ are chosen so that they are as close as possible to the  initial sampling weights $d_k$, according to some distance $\Phi_s$ and subject to some constraints. More exactly, 
\begin{align}
(w_k)_{k\in s}&=\mbox{argmin}_w\Phi_s(w)\label{distance}\\
\mbox{subject to}\quad \sum_{k\in s}w_k\mathbf{x}_k&=t_{\mathbf{x}}, \label{calib_constraint}
\end{align}
where $w = (w_k, k \in s)$ is the vector of weights assigned to each unit in the sample  and $t_{\mathbf{x}}=\sum_{k\in U} \mathbf{x}_k$ is the vector whose elements are the known totals of $\mathcal X_j$ for  $j=1, \ldots, p$. Several distance functions $\Phi_s$ have been studied in Deville and S\"arndal (1992). Under weak regularity assumptions these authors  have shown that all resulting estimators are asymptotically equivalent to the one obtained by minimizing the chi-square distance function  $\Phi_s(w)=\sum_{k\in s}(w_k-d_k)^2/q_kd_k,
\label{chi-square}$
where the $q_k$'s are known positive constants that can be used to take account of the variability of the observations and are unrelated to $d_k$.   A common use in the applications is to consider uniform weights $q_k=1$ for all units $k$ and  we will suppose, without loss of generality, that $q_k=1$ in the following. We will only consider calibration estimators derived using the chi-square distance. The calibration weights $w_k$, $k \in s$, are 
\begin{align}
w_k &=d_k-d_k\mathbf{x}^T_k\left(\sum_{\ell\in s}d_\ell\mathbf{x}_\ell\mathbf{x}^T_\ell\right)^{-1}(\hat{t}_{\mathbf{x}d}-t_{\mathbf{x}}),
\label{calage_DS}
\end{align} 
and the corresponding calibration estimator is obtained by plugging-in $w_k$ in (\ref{def:calibest}).  

With a different point of view, it can  be shown that the  calibration estimator obtained with the chi-squared distance is equal to the generalized regression estimator (GREG) which is derived by assuming a linear regression model between the  study variable $\mathcal{Y}$ and the auxiliary variables $\mathcal{X}_1, \ldots, \mathcal{X}_p$,
\begin{align}
\xi:\quad y_k &=\mathbf{x}^T_k\boldsymbol\beta+\varepsilon_k,
\label{model_xi}
\end{align}
where  $\varepsilon = (\varepsilon_k, k \in U)$ is a  centered random vector with a diagonal variance matrix, whose diagonal elements are equal to $1/q_k$. Cassel \textit{et al.} (1976) suggested the generalized difference estimator 
\begin{eqnarray}
\tilde t_{y,\mathbf{x}}^{\rm{diff}}=\hat{t}_{yd}-\left(\hat{t}_{\mathbf{x}d}-t_{\mathbf{x}}\right)^T\tilde{\boldsymbol{\beta}}_{\mathbf{x}},\label{diff_estim_X}
\end{eqnarray}
where $\tilde{\boldsymbol{\beta}}_{\mathbf{x}}=(\sum_{k\in U}\mathbf{x}_k\mathbf{x}^T_k)^{-1}\sum_{k\in U}\mathbf{x}_ky_k$ is the ordinary least squares estimator of $\boldsymbol{\beta}$ and  $\hat{t}_{\mathbf{x}d}=\sum_{k\in s}d_k\mathbf{x}_k$  is the HT estimator of $t_{\mathbf{x}}$. Remark that  $\tilde t_{y,\mathbf{x}}^{\rm{diff}}$ can not be  computed  because $\tilde{\boldsymbol{\beta}}_{\mathbf{x}}$ can not be  computed unless we have observed the whole population. We estimate $\tilde{\boldsymbol{\beta}}_{\mathbf{x}}$ by $\hat{\boldsymbol{\beta}}_{\mathbf{x}}= \left(\sum_{k\in s}d_k\mathbf{x}_k\mathbf{x}^T_k \right)^{-1}\sum_{k\in s}d_k\mathbf{x}_ky_k$ and obtain the GREG estimator   of $t_y:$ $\hat{t}_{yw}=\hat{t}_{yd}-\left(\hat{t}_{\mathbf{x}d}-t_{\mathbf{x}}\right)^T\hat{\boldsymbol{\beta}}_{\mathbf{x}}.$

Under mild regularity assumptions,  Deville and S\"arndal (1992) have proven that the calibration estimator $\hat t_{yw}$ and $\tilde t_{y,\mathbf{x}}^{\rm{diff}}$ have the same asymptotic distribution. We have $N^{-1} \left(\hat t_{yw}-t_y \right) =N^{-1}\left(\tilde t_{y,\mathbf{x}}^{\rm{diff}}-t_y \right)+o_{\rm p}(n^{-1/2})
\label{approx:tyw}$ and as a result, the asymptotic variance of $\hat t_{yw}$ is  $AV(\hat t_{yw})=\sum_{k\in U}\sum_{\ell\in U}(\pi_{k\ell}-\pi_k\pi_\ell)d_kd_\ell(y_k-\mathbf{x}^T_k\tilde{\boldsymbol{\beta}}_{\mathbf{x}})(y_{\ell}-\mathbf{x}^T_\ell\tilde{\boldsymbol{\beta}}_{\mathbf{x}}),\label{var_calib_est}$
where $\pi_{k\ell} = \rm{Pr}(k\in s \ \& \ \ell \in s)$ is the probability that both $k$ and $\ell$ are included in the sample $s$, and $\pi_{kk}=\pi_k$.
Calibration  will improve the HT estimator, namely $AV(\hat t_{yw})\leq V(\hat t_{yd}),$ if the predicted values $\mathbf{x}^T_k\tilde{\boldsymbol{\beta}}_{\mathbf{x}}$ are close enough to the $y_k$'s,  that is to say if the model $\xi$ stated in (\ref{model_xi}) explains sufficiently well the variable of interest. Nevertheless, when a very large number $p$ of auxiliary variables are used, this result is no longer true as it was remarked by Silva and Skinner (1997)   in a simulation study.
%Usually, with calibration, the user requires that the weight ratios $w_{k}/d_k$ are restricted to lie between pre-specified lower and upper bounds, called also range restrictions. The main reason for doing that is to avoid negative or extremely large weights. Nevertheless, the calibration weights derived for many auxiliary variables may be very unstable and very large, implying that the range restrictions are more difficult to be satisfied. To cope with this issue, several modifications of the distance have been suggested in the literature (Deville and S\"arndal, 1992; Jayasuriya and Valliant, 1996 and Singh and Mohl, 1996), but as Beaumont and Bocci (2008) remarked, these methods ``are all iterative and may not yield a solution even if the range restrictions are mild''. 

%Note also that in the extreme case in which the matrix   $\sum_{k\in s}d_k\textbf{x}_k\textbf{x}^T_k$ is not a full rank matrix, calibration weights  cannot be computed directly with (\ref{calage_DS}) because there is an identifiability issue that is due to multicollinearity and a generalized inverse of previous matrix should be used. Th\'eberge (1999) considered the minimum norm least squares method and the Moore-Penrose inverse matrix to derive weights in presence of multicollinearity among regressors.   

 One way to circumvent the problems due to over-calibration such as extremely large weights and variance inflation, is to relax the calibration constraints, meaning that the too restrictive requirement of being  exactly calibrated  as in (\ref{calib_constraint}) is dropped off and replaced by the requirement of being only approximately calibrated. Then, the deviation  between $\sum_{k\in s}w_k\mathbf{x}_k$ and $\sum_{k\in U}\mathbf{x}_k$ is controlled by means of a penalty. Bardsley and Chambers (1984), in a model-based setting, and Chambers (1996), Rao and Singh (1997) in a design-based setting, suggested finding weights satisfying (\ref{distance}), subject to a quadratic constraint, as %the vector penalized calibration weights $w^{\rm{pen}}$ is obtained  as  the solution of the penalized minimization problem:
$
w^{\rm{pen}}(\lambda)= \mbox{arg} \min_{w}\Phi_s(w)+\lambda^{-1}\left(\hat t_{\mathbf{x}w}-t_{\mathbf{x}}\right)^T \mathbf{C} \left(\hat t_{\mathbf{x}w}-t_{\mathbf{x}}\right),
\label{penal_calib}
$
where $\hat t_{\mathbf{x}w}=\sum_{k\in s}w_k\mathbf{x}_k,$   $\mathbf{C}=\mbox{diag}(c_j)_{j=1}^{p}$, and $c_j\geq 0$ is a user-specified cost associated with the $j$th calibration constraint. The tuning parameter $\lambda >0$ controls the trade off between exact calibration ($\lambda\rightarrow 0$) and no calibration ($\lambda\rightarrow\infty$).  With the chi-square distance, the solution is, for $k \in s$, $w_k^{\rm{pen}}(\lambda)=d_k-d_k\mathbf{x}^T_k\left(\sum_{\ell\in s}d_\ell\mathbf{x}_\ell\mathbf{x}^T_\ell+\lambda\mathbf{C}^{-1}\right)^{-1}\left(\hat{t}_{\mathbf{x}d}-t_{\mathbf{x}} \right),
\label{penal_weights}$
and the penalized calibration estimator is a GREG type estimator, % as given in (\ref{greg}), 
 whose regression coefficient is  estimated by a ridge-type estimator (Hoerl and Kennard, 1970): 
 \[
 \hat{\boldsymbol{\beta}}_{\mathbf{x}}(\lambda) =\left(\sum_{k\in s}d_k\mathbf{x}_k\mathbf{x}^T_k+\lambda\mathbf{C}^{-1}\right)^{-1}\sum_{k\in s}d_k\mathbf{x}_ky_k.
 \]
%Beaumont and Bocci (2008) considered the optimization problem (\ref{penal_calib}) with a general distance. Considering a quadratic penalty in (\ref{penal_calib}) leads to adding  the diagonal matrix $\lambda\mathbf{C}^{-1}$ to   $\sum_{k\in s}d_k\mathbf{x}_k\mathbf{x}^T_k$.  
For an infinite cost $c_j$,  the $j$th calibration constraint is satisfied exactly (see Beaumont and Bocci, 2008). As noted in Bardsley and Chambers (1984), the risk of having negative weights (in the case of the chi-square distance) is greatly reduced by using penalized calibration.  In the context of empirical likelihood approach, Chen \textit{et al.} (2002) suggested  replacing the true totals $t_{\mathbf{x}}$ with $t_{\mathbf{x}}+\Delta(\hat t_{\mathbf{x}w}-t_{\mathbf{x}})$ where $\Delta$ is a diagonal matrix depending on the costs $c_j$ and a tuning parameter controling the deviation between $\hat t_{\mathbf{x}w}$ and $\hat t_{\mathbf{x}}$. 
 Different algorithms have been studied in the literature to select good values, with data driven procedures, of the ridge parameter $\lambda$ (see Bardsley and Chambers, 1984; Chen \textit{et al.} (2002); Beaumont and Bocci, 2008; Guggemos and Till\'e, 2010).

 %Then, the deviation  between $\sum_{k\in s}w_k\mathbf{x}_k$ and $\sum_{k\in U}\mathbf{x}_k$ is controlled by means of a penalty. A class of penalized estimators was suggested by Bardsley and Chambers (1984) in a model-based setting and extended later by Chambers (1996) and by Rao and Singh (1997, 2009) in a design-based (or model-assisted) setting. These approaches lead to a class of model-based or GREG-type estimators that use regression coefficients estimated by ridge-type estimators.
 
%In a design-based setting,%In a calibration approach, the penalized estimator is design-consistent for any fixed value of $\lambda$ and its asymptotic variance is equal to the variance of the generalized difference estimator (\ref{diff_estim_X}) with a ridge-type regression estimator $\tilde{\boldsymbol{\beta}}_{\mathbf{x}}(\lambda) =\left(\sum_{k\in U}\mathbf{x}_k\mathbf{x}^T_k+\lambda\mathbf{C}^{-1}\right)^{-1}\sum_{k\in U}\mathbf{x}_ky_k.$
\par

\vspace{1cm}
\setcounter{chapter}{3}
\setcounter{equation}{0} %-1
\noindent {\bf 3. Calibration on Principal Components}

%\vspace{0.5cm}

We consider in this work another class of approximately  calibrated estimators which are based on dimension reduction through principal components analysis (PCA). In  multivariate statistics  PCA is one of most popular techniques for reducing the dimension of a set of quantitative variables (see {\it e.g.} Jolliffe, 2002) by extracting most of the variability of the data by projection on a low dimension space. %the best known techniques of multivariate analysis who deals with reducing the dimensionality of a over fully abundant data set  while keeping the maximum of variation from the initial data set. 
Principal components analysis consists in transforming the initial data set  into a new set of a few uncorrelated synthetic variables, called principal components (PC), which are linear combinations of the initial variables with the largest variance. The principal components are ``naturally'' ordered, with respect to their contribution  to the total variance of the data, and the reduction of the dimension is then realized by taking only the first few  of PCs. PCA is particularly useful when the correlation among the variables in the dataset is strong.
These new variables can be also used as auxiliary information for calibration as noted in Goga \textit{et al}. (2011) and Shehzad (2012).

% We present  below the method and give some extensions.
%principal component regression (PCR) (Jolliffe, 2002). The PCR calibration consists in reducing the space spanned  by the columns of $\mathbf{X}$ and consider the classical calibration on the new space.

\vspace{0.3cm}
\noindent\textbf{Complete Auxiliary Information }\label{PC_calibration}
%\vspace{0.2cm}

\noindent We suppose without loss of generality that the auxiliary variables are centered, namely  $N^{-1}t_{\mathbf{x}}=0$ and to avoid heavy notations we do not include an intercept term in the model. Note that in applications this intercept term should be included. We suppose now that the auxiliary information is complete, that is to say the $p$-dimensional vector $\mathbf{x}_k$ is known for all the units $k \in U$.
 
Let $\mathbf{X}$ be the $N\times p$ data matrix having $\mathbf{x}^T_k, k\in U$ as rows.  The variance-covariance matrix of the original variables $\mathcal{X}_1, \ldots, \mathcal{X}_p$ is given by $N^{-1}\mathbf{X}^T\mathbf{X}. $ Let $\lambda_1\geq \ldots \geq \lambda_p\geq 0$ be the eigenvalues of  $N^{-1}\mathbf{X}^T\mathbf{X}$ associated to the corresponding orthonormal eigenvectors  $\mathbf{v}_1, \ldots, \mathbf{v}_p$,
\begin{align}
\frac{1}{N}\mathbf{X}^T\mathbf{X}\mathbf{v}_j &=\lambda_j\mathbf{v}_j, \quad j=1,\ldots, p.
\label{eigenelem_pop}
\end{align}
For $j =1, \ldots, p$, the $j$th principal component, denoted by $\mathbf{Z}_j$, is defined as follows
\begin{align}
\mathbf{Z}_j &= \mathbf{X}\mathbf{v}_j =(z_{kj})_{k\in U}.
\label{PC_pop}
\end{align}
Each variable $\mathbf{Z}_j$ has a (population) variance equal to $N^{-1}\sum_{k \in U} z_{kj}^2 = \lambda_j$.
We  only  consider  the first $r$ (with $r<p$) principal components,  $\mathbf{Z}_1,\ldots, \mathbf{Z}_r,$ which correspond to the  $r$ largest eigenvalues.
 In a survey sampling framework, the goal is not to give interpretations of these new variables  $\mathbf{Z}_1,\ldots,\mathbf{Z}_r$ as it is the custom in PCA. These variables serve as a tool to obtain calibration weights which are more stable than the calibration weights that would have been obtained with the whole set of auxiliary variables.

%The method we suggest consist in taking as new calibration variables the PC variables  $\mathbf{Z}_1, \ldots, \mathbf{Z}_r. $ Their population totals is equal to zero since the variables $\mathbf{X}_{j},$ for $j=1\ldots, p$ are considered centered.
%\begin{eqnarray}
%t_{\mathbf{Z}_j}=\sum_Uz_{kj}=0, \quad j=1\ldots, r\label{totaux_pc}
%\end{eqnarray}
%$\boldsymbol{\mathcal{Z}}$ be the population matrix formed by the first $r$ PC, 
%\begin{eqnarray}
%\boldsymbol{\mathcal{Z}}=(\mathbf{Z}_1, \ldots, \mathbf{Z}_r). \label{relation_PC}
%\end{eqnarray}
%Let $\boldsymbol{z}'_k=(z_{k1}, \ldots, z_{kr})$  be the vector containing the values of the $r$ principal components for the $k$-th individual and we may write $\boldsymbol{\mathcal{Z}}=(\boldsymbol{z}'_k)_{k=1}^N.$\\
\noindent More exactly, we want to find the principal component (PC) calibration estimator $\hat{t}_{yw}^{\rm{pc}}(r) = \sum_{k\in s}w_k^{\rm{pc}}(r)y_k,$
%\begin{align*}
%\hat{t}_{yw}^{\rm{pc}}(r) &= \sum_{k\in s}w_k^{\rm{pc}}(r)y_k,
%\end{align*}
where the vector of PC calibration weights $w_k^{\rm{pc}}(r), k\in s$, which depends on the number $r$ of principal components used for calibration, is the solution of the optimization problem (\ref{distance}) and 
subject to
%\begin{align*}
$\sum_{k\in s} w_k^{\rm{pc}}(r) \mathbf{z}_{kr} =   \sum_{k\in U} \mathbf{z}_{kr}$,
%\end{align*}
where $\mathbf{z}^T_{kr}=(z_{k1}, \ldots, z_{kr})$ is the vector containing  the values of the   first  $r$ PCs computed for the $k$-th individual.  The PC calibration weights $w_k^{\rm{pc}}(r)$'s are given by $w_k^{\rm{pc}}(r) =d_k-d_k\mathbf{z}^T_{kr}\left(\sum_{\ell \in s}d_\ell \mathbf{z}_{\ell r}\mathbf{z}^T_{\ell r}\right)^{-1}(\hat{t}_{\mathbf{z}_rd}-t_{\mathbf{z}_r}), k\in s
%\mathbf{d}_s-\mathbf{\tilde{\Pi}}_s^{-1}\boldsymbol{\mathcal{Z}}_{s}\left(\boldsymbol{\mathcal{Z}}'_{s}\mathbf{\tilde{\Pi}}_s^{-1}\boldsymbol{\mathcal{Z}}_{s}\right)^{-1}\left(\mathbf{d}'_s\boldsymbol{\mathcal{Z}}_{s}-\mathbf{1}'_U\boldsymbol{\mathcal{Z}}\right)'.
$
where $\hat{t}_{\mathbf{z}_rd}=\sum_{k\in s}d_k\mathbf{z}_{kr}$ is the HT estimator of the  total $t_{\mathbf{z}_r}= (0, \ldots, 0)$ 
since we have supposed that the original variables have mean zero. % so that the principal components are also centered variables. 

The total $t_y$ is again estimated by a GREG-type estimator which uses $\mathbf{Z}_1, \ldots, \mathbf{Z}_r$ as auxiliary variables 
\begin{align}
\hat{t}_{yw}^{\rm{pc}} (r) &=  \sum_{k\in s}w_k^{\rm{pc}}(r)y_k=\hat{t}_{yd}-\left(\hat{t}_{\mathbf{z}_rd}-t_{\mathbf{z}_r}\right)^T\hat{\boldsymbol{\gamma}}_{\mathbf{z}}(r),
\label{pcestimator}
\end{align}
where 
\begin{align}
\hat{\boldsymbol\gamma}_{\mathbf{z}}(r) &= \left(\sum_{k\in s}d_k\mathbf{z}_{kr}\mathbf{z}^T_{kr} \right)^{-1}\sum_{k\in s}d_k\mathbf{z}_{kr}y_k.
\label{gamma_estim_plan}
\end{align}

The PC calibration estimator $\hat{t}_{yw}^{\rm{pc}}(r)$ depends on the number $r$ of  PC variables and it can be noted  that 
if $r=0$, that is to say if we do not take auxiliary information into account, then $\hat{t}_{yw}^{\rm{pc}}(0)$ is simply the HT estimator (or the H\'ajek estimator if the intercept term is included in the model) whereas if $r=p$, we get the calibration estimator which takes account of all the auxiliary variables.

%\begin{proposition} For $r=0,$ we obtain  the HT estimator,  $\hat{t}_{yw}^{PC}(0)=\hat t_{yd}$ and for $r=p,$ we  obtain the calibration estimator, $\hat{t}_{yw}^{PC}(p)=\hat t_{yw}.$
%\end{proposition}

%We can remark that $\hat{t}_{PC}$ is By its construction we achieve a reduction in dimension of $\mathbf{X}$ by retaining maximum information. Nevertheless, this method demands knowing $\mathbf{X}$ over the whole population in order to derive the eigenvalues and eigenvectors. 
\vspace{0.3cm}
\noindent\textbf{A Model-Assisted Point of View}

Consider again the superpopulation model $\xi$ presented in (\ref{model_xi})  and denote by $\mathbf{G}=(\mathbf{v}_1, \ldots, \mathbf{v}_p)$ the matrix whose $j$th column is the $j$th eigenvector $\mathbf{v}_j$.  Model  $\xi$  may be written in the equivalent form
% $\mathbf{X}\boldsymbol{\beta}=\mathbf{Z}\boldsymbol{\gamma}$ 
\begin{align*}
\xi:\quad y_k &=\mathbf{z}^T_{k}\boldsymbol{\gamma} +\varepsilon_{k},
\end{align*}
where $\boldsymbol{\gamma}=\mathbf{G}^T\boldsymbol{\beta}$ and $\mathbf{z}^T_{k}=(z_{k1}, \ldots, z_{kp})$ where  $z_{kj}$ is the value of $\mathbf{Z}_j$ for the $k$th unit. Principal components regression consists in considering a reduced linear regression model, denoted by $\xi_r$, which uses as predictors the  first $r$ principal components, $\mathbf{Z}_1, \ldots, \mathbf{Z}_r,$ as follows
\begin{align}
\xi_r: \quad y_k &=\mathbf{z}^T_{kr}\boldsymbol{\gamma}(r) +\varepsilon_{kr},
\label{reduced_model}
\end{align}
where $\boldsymbol{\gamma}(r)$ is a vector of $r$ elements composed of the  first $r$ elements of $\boldsymbol{\gamma}$ and $\varepsilon_{kr}$ is the appropriate error term of mean zero. The  least squares estimation, at the population level, of  $\boldsymbol{\gamma}(r)$, is 
\begin{align}
\tilde{\boldsymbol\gamma}_{\mathbf{z}}(r) &= \left(\sum_{k\in U}\mathbf{z}_{kr}\mathbf{z}^T_{kr} \right)^{-1}\sum_{k\in U}\mathbf{z}_{kr}y_k,
\label{tilde_gamma} 
\end{align} 
which in turn can be  estimated, on a sample $s$, by the design-based estimator $\hat{\boldsymbol\gamma}_{\mathbf{z}}(r)$ given by (\ref{gamma_estim_plan}). We can see now that the PC calibration estimator given in (\ref{pcestimator}) is in fact equal to a GREG-type  estimator assisted by the reduced model  $\xi_r$ described in  (\ref{reduced_model}). Note also that since the principal components are centered and uncorrelated, the matrix $\left(\sum_{k\in U}\mathbf{z}_{kr}\mathbf{z}^T_{kr} \right)$ is diagonal, with diagonal elements $( \lambda_1N, \ldots, \lambda_r N)$.

When there is a strong multicollinearity among the auxiliary variables, it is well known that the ordinary least squares estimator of $\boldsymbol\beta$,
%$\tilde{\boldsymbol{\beta}}_{\mathbf{x}}= \left(N^{-1}\sum_{k\in U}\mathbf{x}_k\mathbf{x}^T_k \right)^{-1} N^{-1}\sum_{k\in U}\mathbf{x}_ky_k,$
is very sensitive to small changes in $\mathbf{x}_k$ and $y_k$ and it has a very large variance (see {\textit{e.g} Hoerl and Kennard,~1970). To see better how small eigenvalues may affect $\tilde{\boldsymbol{\beta}}_{\mathbf{x}}$,  Gunst and Mason~(1977) write the least squares estimator as follows:
$\tilde{\boldsymbol{\beta}}_{\mathbf{x}} %&=\left(\sum_{j=1}^p\frac{1}{\lambda_j}\mathbf{v}_j\mathbf{v}_j^T \right) \frac{1}{N}\sum_{k\in U}\mathbf{x}_ky_k\\
= \left(N^{-1}\sum_{k\in U}\mathbf{x}_k\mathbf{x}^T_k \right)^{-1} (N^{-1}\sum_{k\in U}\mathbf{x}_ky_k)=\sum_{j=1}^p\frac{1}{\lambda_j} \left[\mathbf{v}_j^T \left(N^{-1}  \sum_{k \in U} \mathbf{x}_ky_k\right)\right] \mathbf{v}_j.
$
Approximating the covariance matrix $N^{-1}\sum_{k\in U}\mathbf{x}_k\mathbf{x}^T_k=N^{-1}\mathbf{X}^T\mathbf{X}$ by the rank $r$ matrix $\left(\sum_{j=1}^r \lambda_j\mathbf{v}_j\mathbf{v}_j^T \right)$  leads to consider the following approximation to the regression estimator that is based on the  first $r$ principal components,
\begin{align}
\tilde{\boldsymbol\beta}^{\rm{pc}}_{\mathbf{x}}(r)  % &=\mathbf{G}_r\tilde{\boldsymbol{\gamma}}_{\mathbf{z}}(r), \nonumber \\
&= \sum_{j=1}^r\frac{1}{\lambda_j} \left[\mathbf{v}_j^T \left( \frac{1}{N}\sum_{k \in U} \mathbf{x}_ky_k\right)\right] \mathbf{v}_j,  \label{tilde_beta_pc}
\end{align}
where $\mathbf{G}_r=(\mathbf{v}_1, \ldots, \mathbf{v}_r).$ This means that $\tilde{\boldsymbol\beta}^{\rm{pc}}_{\mathbf{x}}(r)$ is obtained by subtracting  from $\tilde{\boldsymbol{\beta}}_{\mathbf{x}}$ the part of the data that belongs to the $p-r$ dimensional space with the smallest variance and by performing the regression in the $r$ dimensional space that contains most of the variability of the data.  Note that ridge-regression (Hoerl and Kennard, 1970),  which is  an alternative way of dealing  with the multicollinearity issue, consists in adding a  positive term $\lambda$ to all eigenvalues $\lambda_j, j=1, \ldots, p$. More exactly, the ridge estimator of $\boldsymbol{\beta}$ may be written as $\tilde{\boldsymbol{\beta}}_{\mathbf{x}}(\lambda)=(N^{-1}\sum_{k\in U}\mathbf{x}_k\mathbf{x}^T_k+\lambda\mathbf{I}_p)^{-1}\left( N^{-1}\sum_{k \in U} \mathbf{x}_ky_k\right)=\sum_{j=1}^p\frac{1}{\lambda+\lambda_j} \left[\mathbf{v}_j^T \left( N^{-1}\sum_{k \in U} \mathbf{x}_ky_k\right)\right] \mathbf{v}_j,$ where $\mathbf{I}_p$ is the $p$-dimensional identity matrix. Both the ridge regression estimator $\tilde{\boldsymbol{\beta}}_{\mathbf{x}}(\lambda)$ and the  principal components estimator $\tilde{\boldsymbol\beta}^{\rm{pc}}_{\mathbf{x}}(r)$ are biased for $\boldsymbol{\beta}$ under the model $\xi$ (Gunst and Mason,~1977).
%\begin{eqnarray*}
%\mbox{Bias}_{\xi}(\tilde{\boldsymbol\beta}_{PC,r}) & = & \mathbf{G}_r\mathbf{G}_r^T\boldsymbol{\beta}-\boldsymbol{\beta}
% =-\sum_{j=r+1}^p\mathbf{v}_j\mathbf{v}^T_j\boldsymbol{\beta}.
%\end{eqnarray*}
%Gunst and Mason (1976) show that  a necessary condition  for $\tilde{\boldsymbol\beta}_{PC,r}$ to be better than $\hat{\boldsymbol\beta}_{OLS}$ from the $\xi$-mean squared error point of view, i.e. $E_{\xi}(\hat{\boldsymbol{\beta}}-\boldsymbol{\beta})(\hat{\boldsymbol{\beta}}-\boldsymbol{\beta})',$ is that 
%$$
%\sum_{j=1}^r\lambda_j(\mathbf{v}'_j\boldsymbol{\beta})^2/\sigma^2<1
%$$ 

The PC regression estimator $\tilde{\boldsymbol\beta}^{\rm{pc}}_{\mathbf{x}}(r)$ can be estimated under the sampling design by 
\begin{align}
\hat{\boldsymbol\beta}^{\rm{pc}}_{\mathbf{x}}(r) &=\mathbf{G}_r\hat{\boldsymbol{\gamma}}_{\mathbf{z}}(r),
\label{estim_betapc_plan}
\end{align}
where $\hat{\boldsymbol{\gamma}}_{\mathbf{z}}(r)$ is given in (\ref{gamma_estim_plan}). Using relation (\ref{estim_betapc_plan}) and the fact that $\mathbf Z_j=\mathbf{X}\mathbf{v}_j,$ we obtain that $\left(\hat{t}_{\mathbf{z}_rd}-t_{\mathbf{z}_r}\right)^T\hat{\boldsymbol{\gamma}}_{\mathbf{z}}(r)=\left(\hat{t}_{\mathbf{x}d}-t_{\mathbf x}\right)^T\hat{\boldsymbol{\beta}}^{\rm{pc}}_{\mathbf{x}}(r).$ Consequently $\hat t^{\rm{pc}}_{yw}(r)$ can also be written as follows,
%\begin{align*}
$
\hat t^{\rm{pc}}_{yw}(r) =  \hat{t}_{yd}-\left(\hat{t}_{\mathbf{x}d}-t_{\mathbf x}\right)^T\hat{\boldsymbol{\beta}}^{\rm{pc}}_{\mathbf{x}}(r),
$
%\end{align*}
and $\hat t^{\rm{pc}}_{yw}(r)$ may be seen as a GREG-type estimator assisted by the model $\xi$ when $\boldsymbol{\beta}$ is estimated by $\hat{\boldsymbol{\beta}}^{\rm{pc}}_{\mathbf{x}}(r)$. 

\vspace{0.3cm}
\noindent\textbf{Calibration on the second moment of the PC variables}

With complete auxiliary information, S\"{a}rndal (2007) stated that ``we are invited to consider $x^2_{kj}, j=1, \ldots, p$ and other functions of $x^2_{kj}$ for inclusion in $\mathbf{x}_k$'' especially when ``the relationship to the study variable is curved''.  Calibration on higher-order moments of the auxiliary variables has also been studied by Ren~(2000). In our case,  the PC variables $\mathbf Z_j$ satisfy $N^{-1}\mathbf{Z}^T_j\mathbf{Z}_j=N^{-1}\sum_{k\in U}z_{kj}^2=\lambda_j, \quad \mbox{for all}\quad j=1, \ldots, p.$  
This means that in presence of complete auxiliary information, the totals of squares of the PCs are known.  As a consequence, if we keep the first $r$ variables $\mathbf{Z}_1, \ldots, \mathbf{Z}_r$ corresponding to the largest $r$ eigenvalues, we can consider $r$ additional  calibration constraints on the second moment of these PCs.   We look for the calibration weights $w^{\rm{pc}}(r)$ solution to  (\ref{distance}) and subject to $\sum_{k\in s} w_k^{\rm{pc}}(r) \left(\mathbf{z}_{kr},\mathbf{z}^2_{kr}\right)^T = \sum_{k\in U} \left(\mathbf{z}_{kr},\mathbf{z}^2_{kr}\right)^T$
where $\mathbf{z}^2_{kr}=(z^2_{k1}, \ldots, z^2_{kr})$. 
%The calibration estimator  is obtained easily  in fact a generalized regression estimator for the auxiliary information $\mathbf{T}$ of dimension $N\times (2r),$  as follows

The estimator derived in this way is expected to perform better than the estimator calibrated only on the first moment of the principal components. Nevertheless, calibration on the second moment of the PCs requires $r$ additional calibration constraints.

%\vspace{0.3cm}
%\noindent\textbf{Calibration on Estimated Principal Components}\label{EPC}

\vspace{1cm}
\setcounter{chapter}{4}
\setcounter{equation}{0} %-1
\noindent {\bf 4. Calibration on Estimated Principal Components}

\vspace{0.5cm}

The approach presented in the above Sections  supposes that the values of the auxiliary variables $\mathcal{X}_j,$ for $j=1,\ldots, p$ are known for all units $k$ in the population $U$. In practice, it often  happens that the variables $\mathcal{X}_j$ are  only known for the sampled individuals, but their population totals are known. Then, it is not possible anymore to compute  the eigenvalues and the eigenvectors of the population variance-covariance matrix. We present in this section a way to perform principal components calibration  when the auxiliary variables are only observed for the units belonging to the sample.\\
Let $\boldsymbol{\Gamma}=N^{-1}\mathbf{X}^T\mathbf{X}$ be  the variance-covariance matrix estimated by 
\begin{align}
\hat{\boldsymbol{\Gamma}}&=\frac{1}{\hat N}\sum_{k\in s}d_k(\mathbf{x}_k-\hat{\overline{\mathbf{X}}})(\mathbf{x}_k-\hat{\overline{\mathbf{X}}})^T=\frac{1}{\hat N}\sum_{k\in s}d_k\mathbf{x}_k\mathbf{x}^T_k-\hat{\overline{\mathbf{X}}}\hat{\overline{\mathbf{X}}}^T,
\label{estim_gamma}
\end{align}
where $\hat N=\sum_{k\in s}d_k$ and $\hat{\overline{\mathbf{X}}}=\hat N^{-1}\sum_{k\in s}d_k\mathbf{x}_k$. Let $\hat{\lambda}_1\geq \ldots \geq \hat{\lambda}_p \geq 0$  be the sorted eigenvalues  of $\hat{\boldsymbol{\Gamma}}$ and  $\hat{\mathbf{v}}_1, \ldots, \hat{\mathbf{v}}_p$ the corresponding orthonormal eigenvectors, 
\begin{align}
\hat{\boldsymbol{\Gamma}} \hat{\mathbf{v}}_j &= \hat{\lambda}_j \hat{\mathbf{v}}_j, \quad j=1,\ldots, p.
\label{eigenelem_ech}
\end{align}
We have that $\hat{\lambda}_j$ and $\hat{\mathbf{v}}_j$ are the design-based estimators of $\lambda_j$ and respectively, $\mathbf{v}_j$ for  $j=1, \ldots, p$. It is shown in  Cardot \textit{et al.} (2010)  that with large samples and under classical assumptions on  the first and second order inclusion probabilities $\pi_k,$ $\pi_{kl}$ as well as on the variables  $\mathcal X_j,$ (see the assumptions (A1)-(A6) in Section~5), that the estimators $\hat{\lambda}_j$ and $\hat{\mathbf{v}}_j$ are asymptotically design unbiased and consistent for $\lambda_j$ and respectively, for $\mathbf{v}_j$. 
%More precisely,
%\begin{align*}
%$\rm{E}_p\left( \max_j|\hat{\lambda}_j-\lambda_j| \right)^2 =O(n^{-1})$, %\quad \mbox{and}
%\end{align*}
%where $\rm{E}_p(.)$ denotes the expectation with respect to the sampling design. If $\lambda_{j-1}>\lambda_j > \lambda_{j+1}$, it can also be shown that, for $j=1, \ldots, p$, 
%\begin{align*}
%$\rm{E}_p \left( \left\| \hat{\mathbf{v}}_j-\mathbf{v}_j \right\|^2\right) =O(n^{-1})$,
%\end{align*}
%where $\| \cdot \|$ is the usual Euclidean norm and $\hat{\mathbf{v}}_j$, which is defined up to sign change, is chosen to satisfy $\hat{\mathbf{v}}_j^T \mathbf{v}_j \geq 0$. %$\frac{1}{N}\mathbf{X}'_s\boldsymbol{\Pi}^{-1}_s\mathbf{X}_s $

The unknown population principal components $\mathbf{Z}_j$ defined in (\ref{PC_pop}) can be approximated  as follows, 
%\begin{align*}
$\hat{\mathbf{Z}}_j = \mathbf{X}\hat{\mathbf{v}}_j,$
%\end{align*}
%We have $\hat{\mathbf{Z}}_j=(\hat z_{kj})_{k\in U}$ with $\hat z_{kj},$ for all $k\in U$  estimating  the elements $z_{kj}$ of  $\mathbf{Z}_j. $ 
reminding  that $\hat{\mathbf{Z}}_j=(\hat z_{kj})_{k\in U}$ is  only  known for the units in the sample. Nevertheless, its population total  $t_{\hat{\mathbf{Z}}_j}=\sum_{k\in U}\hat z_{kj}$ is known and is equal to zero since $t_{\hat{\mathbf{Z}}_j} =t^T_{\mathbf{x}}\hat{\mathbf{v}}_j=0, \quad j=1, \ldots, p.$
Note also that $\hat{\mathbf{Z}}_j$ are not exactly the principal components associated with the variance-covariance matrix $\hat{\boldsymbol{\Gamma}}$ because the original variables are  centered in the population but not necessarily  in the sample.

 Consider now the first  $r$ estimated principal components
$\hat{\mathbf{Z}}_1, \ldots, \hat{\mathbf{Z}}_r$,
corresponding to the  $r$ largest eigenvalues $\hat{\lambda}_1 \geq \ldots \geq \hat{\lambda}_r \geq 0$ and suppose that $\hat{\lambda}_r>0$.
%Remark that the number of PC considered here may be different from the one considered in section \ref{PC_calibration}  but, for ease of notation, we will use the same $r.$  
The estimated principal component (EPC) calibration estimator of $t_y$ is 
%\begin{align*}
$\hat t_{yw}^{\rm{epc}}(r) =\sum_{k\in s}w_k^{\rm{epc}}(r)y_k$,
%\end{align*}
where the EPC calibration weights $w_k^{\rm{epc}}, k \in s$ are the solution of the optimization problem (\ref{distance}) subject to the constraints
$\sum_{k\in s} w_k^{\rm{epc}}(r) \mathbf{\hat{z}}_{kr} =   \sum_{k\in U} \mathbf{\hat z}_{kr},$
where $\mathbf{\hat{z}}^T_{kr}=(\hat{z}_{k1}, \ldots, \hat{z}_{kr})$ is the  vector of values of $\hat{\mathbf{Z}}_j,$ $j=1, \ldots, r$ recorded for the $k$th unit. 
 With the chi-square distance function~$\Phi_s$, the EPC calibration weights $w_k^{\rm{epc}}(r)$ are given by 
\begin{align}
w_k^{\rm{epc}}(r) & =d_k-d_k\mathbf{\hat z}^T_{kr}\left(\sum_{\ell \in s}d_\ell \mathbf{\hat z}_{\ell r}\mathbf{\hat z}^T_{\ell r}\right)^{-1}(\hat{t}_{\mathbf{\hat z}_rd}-t_{\mathbf{\hat z}_r}),
%\mathbf{d}_s-\mathbf{\tilde{\Pi}}_s^{-1}\boldsymbol{\mathcal{Z}}_{s}\left(\boldsymbol{\mathcal{Z}}'_{s}\mathbf{\tilde{\Pi}}_s^{-1}\boldsymbol{\mathcal{Z}}_{s}\right)^{-1}\left(\mathbf{d}'_s\boldsymbol{\mathcal{Z}}_{s}-\mathbf{1}'_U\boldsymbol{\mathcal{Z}}\right)'.
\label{hatpcestimator}
\end{align}
where $\hat{t}_{\mathbf{\hat z}_rd}=\sum_{k\in s}d_k\mathbf{\hat z}_{kr}$ is the HT estimator of the total $t_{\mathbf{\hat z}_r}=\sum_{k\in U}\mathbf{\hat z}_{kr}=0.$ The EPC calibration estimator for $t_y$ is given by
%\begin{align}
$
\hat t_{yw}^{\rm{epc}}(r) = \sum_{k\in U}w_k^{\rm{epc}}(r)y_k= \hat{t}_{yd}-\left(\hat{t}_{\mathbf{\hat{z}}d}-t_{\mathbf{\hat{z}}}\right)^T\hat{\boldsymbol{\gamma}}_{\mathbf{\hat{z}}}(r),$
%\end{align}
where $\hat{\boldsymbol{\gamma}}_{\mathbf{\hat{z}}}(r)=\left(\sum_{k\in s}d_k\mathbf{\hat z}_{kr}\mathbf{\hat z}^T_{kr}\right)^{-1}\sum_{k\in s}d_k\mathbf{\hat z}_{kr}y_k.$
The EPC calibration estimator may also be written with respect to the population totals of the original variables, $\mathcal X_1, \ldots, \mathcal X_p,$ as follows
% \begin{eqnarray}
$\hat t_{yw}^{\rm{epc}} (r) =  \hat{t}_{yd}-\left(\hat{t}_{\mathbf{x}d}-t_{\mathbf x}\right)^T\hat{\boldsymbol{\beta}}^{\rm{epc}}_{\mathbf{x}}(r)$,
%\end{eqnarray}
where 
$
\hat{\boldsymbol{\beta}}^{\rm{epc}}_{\mathbf{x}}(r)=(\hat{\mathbf{v}}_1, \ldots, \hat{\mathbf{v}}_r)\hat{\boldsymbol{\gamma}}_{\mathbf{\hat{z}}}(r).
$

%%%%%%%%%%%%%%%%%%%%%%%%%%%%%%%%%%%%%%%%%%%%%%
%%%%%%%%%%%%%%%%%%%%%%%%%%%%%%%%%%%%%%%%%%%%%%
%\vspace{1cm}
%\noindent\textbf{Some Asymptotic Properties of the Principal Components Calibration Estimators}\label{subsec:asymptot}
\vspace{1cm}
\setcounter{chapter}{5}
\setcounter{equation}{0} %-1
\noindent {\bf 5. Some Asymptotic Properties of the Principal Components Calibration Estimators}

\vspace{0.5cm}

%%%%%%%%%%%%%%%%%%%%%%%%%%%%%%%%%%%%%%%%%%%%%%

We adopt in this section the asymptotic framework of Isaki and Fuller~(1982). We consider a sequence of growing and nested populations $U_N$ with size $N$ tending to infinity and a sequence of samples $s_N$ of size $n_N$ drawn from $U_N$ according to the fixed-size sampling designs $p_N(s_N)$. The sequence of subpopulations is an increasing nested one, whereas the sample sequence is not. For simplicity of notation, we drop the subscript $N$ in the following when there is no ambiguity. The number $p_N$ of auxiliary variables as well as the number $r_N$ of principal components are allowed to tend to infinity.  %The proofs are given in the Appendix. 
We suppose that the following assumptions hold.

\begin{itemize}
\item[(A1)] $\lim_{N\rightarrow\infty}\frac{n}{N}=\pi\in(0,1).$
\item[(A2)] $\pi_k>\delta>0$ for all $k\in U_N;$
 $\overline{\lim}_{N\rightarrow\infty}n\max_{k\neq l}|\pi_{kl}-\pi_k\pi_l|<\infty.$
\item[(A3)] There is a constant $C_y$ such that for all $N$, $\frac{1}{N}\sum_{U_N} y^4_k< C_y$.
\item[(A4)] %There is a constant $C_x>0$ such that $\| \mathbf{x}_k \|^2 \leq C \ p_N$, for all $k\in U_N,$ where $\| \cdot \|$ is the Euclidean norm.
The largest eigenvalue $\lambda_{1N}$ of $\boldsymbol{\Gamma}_N$ is bounded, $\lambda_{1N} \leq C_\lambda$.  
\item[(A5)] There is a contant $c>0$ and a non decreasing sequence of integers $(r_N)$ such that for all $N \geq N_0$ we have  $\lambda_{r_N} \geq c$.
\item[(A6)] There is a constant $C_4$ such that, $\forall \mathbf{v} \in \mathbb{R}^{p_N}$ satisfying $\| \mathbf{v} \| =1$, we have ${N}^{-1}\sum_{k \in U_N} \left| \langle \mathbf{x}_k, \mathbf{v} \rangle \right|^4 \leq C_4$.
\end{itemize}

Conditions (A1), (A2) and (A3) are classical hypotheses for asymptotics in survey sampling (see {\it e.g.} Breidt and Opsomer, 2000). 
Condition (A4) is closely related to a moment condition on $\mathbf{x}_k,$  for $k \in U_N$. If (A4) is fulfilled, $\frac{1}{N} \sum_{k \in U_N} \left\| \mathbf{x}_k \right\|^2 = \sum_{j=1}^{p_N} \lambda_{jN} \leq C_\lambda p_N$.
Assumption (A5) ensures that there is no identifiability issue for the sequence $\tilde{\boldsymbol\beta}^{\rm{pc}}_{\mathbf{x}}(r_N)$ of regression coefficients defined at the population level. It only deals with $\lambda_{r_N}$ and does not prevent $\lambda_{p_N}$ from  being  equal to zero or from being very small. The finite population moment assumptions (A4) and (A6) indicate that the vectors $\mathbf{x}_k$ cannot be too concentrated in one direction (see Vershynin (2012) for examples in a classical statistical inference context). The proofs of Proposition \ref{result_pc_asympt} and Proposition \ref{result_pc_estim}  are given in a Supplementary file.

We first show that the estimator based on the true principal components  is consistent and we give its asymptotic variance. Note that the assumption on the eigenvalues  $\lambda_r > \lambda_{r+1} \geq 0$ ensures that there is no identifiability problem of the eigenspace generated by the eigenvectors associated to the $r$ largest eigenvalues. 
The condition $r_N^3/n \to 0$ prevents the number of principal components from being too large and ensures that the remainder term, whose order is $r_N^{3/2}/n$, tends to zero and is negligible compared to the main term whose order is $1/\sqrt{n}$. %$\lambda_1 \geq \cdots \geq \lambda_r$.
%very large but is consider fixed. 

%The generalized difference estimator is given by
%\begin{eqnarray}
%\tilde t_{y,\mathbf{x}}^{\rm{diff}}(r) &= &\hat{t}_{yd}-\left(\hat{t}_{\mathbf{x}d}-t_{\mathbf{x}}\right)^T\tilde{\boldsymbol{\beta}}_{PC,r}\\
%& = & \hat{t}_{yd}-\left(\hat{t}_{\mathbf{z}_rd}-t_{\mathbf{z}_r}\right)^T\tilde{\boldsymbol{\gamma}}_{Z_r},
%\end{eqnarray}
%by using relations (\ref{PC_pop}) and (\ref{tilde_beta_pc}).

\begin{result} \label{result_pc_asympt} Assume that (A1)-(A6) hold and that $\lambda_r > \lambda_{r+1} \geq 0$. 
If  $r_N^3/n \to 0$ when $N$ goes to infinity, then 
%Then,  $\hat{\boldsymbol{\gamma}}_{\mathbf{z}}(r)-\tilde{\boldsymbol{\gamma}}_{\mathbf{z}}(r)=O_p(n^{-1/2}) $
\[ N^{-1}(\hat t_{yw}^{\rm{pc}}(r_N)-t_y) =N^{-1} \left(\tilde t_{y,\mathbf{x}}^{\rm{diff}}(r_N)-t_y \right)+O_p \left(\frac{r_N^{3/2}}{n} \right)
\]
and  $\tilde t_{y,\mathbf{x}}^{\rm{diff}}(r_N)=\hat{t}_{yd}-\left(\hat{t}_{\mathbf{x}d}-t_{\mathbf{x}}\right)^T\tilde{\boldsymbol{\beta}}^{\rm{pc}}_{\mathbf{x}}(r_N)$ satisfies
%\[ 
$N^{-1} \left(\tilde t_{y,\mathbf{x}}^{\rm{diff}}(r_N) -  t_y \right) = O_p \left( \frac{1}{\sqrt{n}} \right)$.
%\]
\end{result} 

The condition  $r_N^3/n \to 0$  could certainly be relaxed for particular sampling designs with high entropy under additional moment assumptions.
Note also that the asymptotic variance of $\hat t_{yw}^{\rm{pc}}(r)$ is given by $AV(\hat t_{yw}^{\rm{pc}}(r)) =\sum_{k\in U}\sum_{l \in U}(\pi_{kl}-\pi_k\pi_l) \left(y_k-\mathbf{x}^T_{k}\tilde{\boldsymbol{\beta}}^{\rm{pc}}_{\mathbf{x}}(r) \right) \left(y_l-\mathbf{x}^T_{l}\tilde{\boldsymbol{\beta}}^{\rm{pc}}_{\mathbf{x}}(r) \right).$
%The asymptotic variance may also be written with respect to the principal components $\mathbf{Z}_j$, 
%\begin{align}
%AV(\hat t_{yw}^{\rm{pc}}(r)) &=\sum_{k\in U}\sum_{\ell \in U} \left(\pi_{kl}-\pi_k\pi_l)d_kd_l(y_k-\mathbf{z}^T_{kr}\tilde{\boldsymbol{\gamma}}_{\mathbf{z}}(r) \right) \left(y_l-\mathbf{z}^T_{lr}\tilde{\boldsymbol{\gamma}}_{\mathbf{z}}(r)\right).
%\label{var_PCcalib_est}
%\end{align}
%Thus, the variance can be estimated as follows,
%\begin{align*}
%\widehat{Var}(\hat t_{yw}^{\rm{pc}}(r)) &=\sum_{k\in s}\sum_{l \in s}\frac{\pi_{kl}-\pi_k\pi_l}{\pi_{kl}}d_kd_l \left(y_k-\mathbf{z}^T_{kr}\hat{\boldsymbol{\gamma}}_{\mathbf{z}}(r) \right) \left(y_l-\mathbf{z}^T_{lr}\hat{\boldsymbol{\gamma}}_{\mathbf{z}}(r)\right).
%\end{align*}

%\begin{result} Under the assumptions ???, the variance estimator $\widehat{Var}(\hat t_{yw}^{PC}(r))$ is consistent for the asymptotic variance $AV(\hat t_{yw}^{PC}(r)),$ in the sense that 
%\begin{align*}
%\frac{n}{N^2}(\widehat{Var}(\hat t_{yw}^{PC}(r))-AV(\hat t_{yw}^{PC}(r))) &=o_{\rm{p}}(1).
%\end{align*}
%\end{result} 
%Theoretical properties similar to GREG estimator can be formed such as variance, MSE and unbiasedness under design.\\
%\subsubsection{General distance function}

%We present now an intermediate result which states that we consistently estimate the  regression coefficient based on the estimated principal components.
Before stating a consistency result for calibration on estimated principal components, let us introduce an additional condition on the spacing between adjacent eigenvalues.
\begin{itemize}
\item[(A7)] There is a constant $c_\lambda>0$ such that $\min_{j=1, \ldots, r_N+1}( \lambda_{jN}-\lambda_{j+1,N} ) \geq c_\lambda r_N$.
\end{itemize} 
This assumption ensures that the $r_N$ largest eigenvalues are nearly equidistributed in $[c, C_\lambda]$.

\begin{result}\label{result_pc_estim} 
Assume that  (A1)-(A7) hold. If  $p_N^3 r_N^{3}/n \to 0$ when $N$ goes to infinity, then 
\[
N^{-1}(\hat t_{yw}^{\rm{epc}}(r_N)-t_y) =N^{-1} \left(\tilde{t}_{y,\mathbf{x}}^{\rm{diff}}(r_N)-t_y \right)+O_p \left(\frac{ p_N^{3/2} r_N^{3/2}}{n} \right)
\]
%and $\hat t_{y,\mathbf{x}}^{\rm{diff}}(r_N)  = \hat{t}_{yd}-\left(\hat{t}_{\hat{\mathbf{z}}_rd}-t_{\hat{\mathbf{z}}_r}\right)^T \tilde{\boldsymbol{\gamma}}_{\mathbf{z}}(r_N)$ satisfies
%\[
% \hat{t}_{yd}-\left(\hat{t}_{\hat{\mathbf{z}}_rd}-t_{\hat{\mathbf{z}}_r}\right)^T \tilde{\boldsymbol{\gamma}}_{\mathbf{z}}(r_N) = O_p \left( \sqrt{\frac{r_N}{n}} \right).
%\]

\end{result}
A more restrictive condition on how $r_N$ may go to infinity is imposed when the principal components are estimated.  The condition $p_N^3 r_N^{3}/n \to 0$ ensures that the remainder term of order $p_N^{3/2} r_N^{3/2}/n$ is negligible compared to $1/\sqrt{n}$. 
 If $p_N$ is bounded, one gets back to classical $\sqrt{n}$-rates of convergence whether the population principal components are known or not. 
%The last result simply shows that estimating the principal components does not change the asymptotic behavior of the principal component calibration estimator. 
%\begin{result}\label{result:tepc}
%Assume that  (A1)-(A4) hold and that $\lambda_r > \lambda_{r+1} \geq0$. We have $N^{-1}(\hat t_{yw}^{\rm{epc}}(r)-t_y) =N^{-1} \left(\tilde t_{y,\mathbf{x}}^{\rm{diff}}(r)-t_y \right)+o_p(n^{-1/2}).$
%\end{result}
%The asymptotic variance of $\hat t_{yw}^{\rm{epc}}(r)$ is the variance of $\tilde t_{y,\mathbf{x}}^{\rm{diff}}(r)$ and given in Proposition \ref{result_pc_asympt}. 

If all the second-order inclusion probabilities $\pi_{k\ell}$ are strictly positive, the asymptotic variance of $\hat t_{yw}^{\rm{pc}}(r)$ can be estimated by the Horvitz-Thompson variance estimator for the  residuals $y_k-\mathbf{x}^T_{k}\hat{\boldsymbol{\beta}}^{\rm{pc}}_{\mathbf{x}}(r), k\in s$,
\begin{align*}
\widehat{Var}(\hat t_{yw}^{\rm{pc}}(r)) &=\sum_{k\in s}\sum_{\ell \in s}\frac{\pi_{k\ell}-\pi_k\pi_\ell}{\pi_{k\ell}}d_kd_\ell \left(y_k-\mathbf{x}^T_{k}\hat{\boldsymbol{\beta}}^{\rm{pc}}_{\mathbf{x}}(r) \right) \left(y_\ell-\mathbf{x}^T_{\ell}\hat{\boldsymbol{\beta}}^{\rm{pc}}_{\mathbf{x}}(r) \right),
\end{align*}
while the asymptotic variance of $\hat t_{yw}^{\rm{epc}}(r)$ can be estimated by the Horvitz-Thompson variance estimator for the  residuals $y_k-\mathbf{x}^T_{k}\hat{\boldsymbol{\beta}}^{\rm{epc}}_{\mathbf{x}}(r), k\in s:$
\begin{align*}
\widehat{Var}(\hat t_{yw}^{\rm{epc}}(r)) &=  \sum_{k\in s}\sum_{\ell \in s}\frac{\pi_{kl}-\pi_k\pi_l}{\pi_{kl}}d_kd_\ell (y_k-\mathbf{x}^T_{k}\hat{\boldsymbol{\beta}}^{\rm{epc}}_{\mathbf{x}}(r))(y_\ell-\mathbf{x}^T_{\ell}\hat{\boldsymbol{\beta}}^{\rm{epc}}_{\mathbf{x}}(r)).
\end{align*}

\vspace{0.5cm}
\setcounter{chapter}{6}
\setcounter{equation}{0} %-1
\noindent {\bf 6. Partial Calibration on Principal Components}

\vspace{0.5cm}

The calibration estimators derived before are not designed  to give the exact finite population totals of the original variables $\mathcal{X}_j,$ $j=1, \ldots, p$. In practice, it is often desired to have this property satisfied for  a few but important socio-demographical variables such as sex, age  or  socio-professional category.

We can adapt the method presented in the previous section in order to fulfill this requirement. We split the auxiliary matrix $\mathbf X$ into two blocks:  a first block $ \tilde{\mathbf X}_1$ containing the $p_1$ most important variables with $p_1$ small compared to $p$, and a second block $\tilde{\mathbf X}_2$ containing the remaining $p_2=p-p_1$ variables.  We have 
$\mathbf X=(\tilde{\mathbf X}_1, \tilde{\mathbf X}_2)$. Note that the constant term will generally belong to the first block of variables.

The goal is to get calibration weights such that the totals of the $p_1$ auxiliary variables in $\tilde{\mathbf X}_1$ are estimated exactly while the totals of the $p_2$ remaining variables are estimated only approximately.  The idea is to calibrate directly on the auxiliary variables from $\tilde{\mathbf X}_1$ and on the first principal components of $\tilde{\mathbf X}_2$, after having taken into account the fact that the variables in $\tilde{\mathbf X}_1$ and all their linear combinations  are perfectly estimated. 
For that, we introduce the matrix $\mathbf{I}_{N}$ which is the $N$-dimensional identity matrix and $\mathbf{P}_{\tilde{\mathbf X}_1}=\tilde{\mathbf X}_1(\tilde{\mathbf X}_1^T\tilde{\mathbf X}_1)^{-1}\tilde{\mathbf X}_1^T$ the orthogonal projection onto the vector space spanned by the column vectors of matrix $\tilde{\mathbf X}_1$. We also define the matrix % $\mathbf{A}$,
%\begin{align*}
$\mathbf{A} = \left(\mathbf{I}_{N}-\mathbf{P}_{\tilde{\mathbf X}_1} \right)\tilde{\mathbf X}_2$,
%\end{align*}
which is the projection of $\tilde{\mathbf X}_2$ onto the orthogonal space spanned by the column vectors of $\tilde{\mathbf X}_1$. Matrix $\mathbf{A}$ represents the residual part of $\tilde{\mathbf X}_2$ that is not ``calibrated'' when considering an estimator of the total of $\mathcal{Y}$ calibrated on the variables in $\tilde{\mathbf X}_1$.
We define the residual covariance matrix $N^{-1}\mathbf{A}^T\mathbf{A} =N^{-1}\tilde{\mathbf X}_2^T \left(\mathbf{I}_{N}-\mathbf{P}_{\tilde{\mathbf X}_1} \right)\tilde{\mathbf X}_2$
 and denote by  $\tilde{\lambda}_1\geq\ldots\tilde{\lambda}_{p_2}$ its  eigenvalues and by $\tilde{\mathbf{v}}_1, \ldots, \tilde{\mathbf{v}}_{p_2}$ the 
 corresponding orthonormal eigenvectors. 
Consider now,  
%\begin{align*}
$\tilde{\mathbf{Z}}_j =\mathbf{A}\tilde{\mathbf{v}}_j$, for $j=1, \ldots, p_2$,
%\end{align*}
the principal components of $\mathbf{A}$. The calibration variables are $(\tilde{\mathbf X}_1, \tilde{\mathbf{Z}}_1, \ldots, \tilde{\mathbf{Z}}_r)$ of zero totals and the partial principal component (PPC) calibration estimator of $t_y$ is 
%\begin{align*}
$\hat t_{yw}^{\rm{ppc}}(r) =\sum_{k\in s}w_k^{\rm{ppc}}(r)y_k$,
%\end{align*}
where the PPC calibration weights $w_k^{\rm{ppc}}(r)$, for $k \in s$, are the solution of the optimization problem (\ref{distance}) subject to
$\sum_{k\in s} w_k^{\rm{ppc}}(r)\left(\tilde{\mathbf{x}}_k,\tilde{\mathbf{z}}_{kr}\right)^T=   \sum_{k\in U} \left(\tilde{\mathbf{x}}_k,\tilde{\mathbf{z}}_{kr}\right)^T,$
where $\tilde{\mathbf{x}}_k=(\tilde{x}_{k1}, \ldots, \tilde{x}_{kp_1})$ is the vector of the values of the variables  in $\tilde{\mathbf{X}}_1$ and $\mathbf{z}^T_{kr}=(\tilde{z}_{k1}, \ldots, \tilde{z}_{kr})$ is the vector whose elements are the partial principal components $\tilde{\mathbf{Z}}_1, \ldots, \tilde{\mathbf{Z}}_r$  for unit $k$.
Note that with a different point of view, Breidt and Chauvet (2012) use,  at the sampling stage, similar ideas in order to perform penalized balanced sampling.

Suppose now that we only have a sample $s$ at hand and we know the totals of all  the calibration variables. We denote by $\tilde{\mathbf{X}}_{s,1}$ (resp. $\tilde{\mathbf{X}}_{s,2}$) the $n\times p_1$ (resp. $n \times p_2$) matrix containing the observed values of the auxiliary variables. We can estimate the matrix $\mathbf{A}$ by 
%\begin{align*}
$\hat{\mathbf{A}} = \left(\mathbf{I}_n - \hat{\mathbf{P}}_{\tilde{\mathbf{X}}_{s,1}}\right) \tilde{\mathbf{X}}_{s,2}$
%\end{align*} 
where $\hat{\mathbf{P}}_{\tilde{\mathbf{X}}_{s,1}} = \tilde{\mathbf{X}}_{s,1} \left( \tilde{\mathbf{X}}_{s,1}^T \mathbf{D}_s \tilde{\mathbf{X}}_{s,1} \right)^{-1} \tilde{\mathbf{X}}_{s,1}^T\mathbf{D}_s$, is the estimation of the projection onto the space generated by the columns of $\tilde{\mathbf{X}}_1$ and $\mathbf{D}_s$ is the $n \times n$ diagonal matrix, with diagonal elements $d_k, \ k \in s$. Then, we can perform the principal components analysis of  the projected  sampled data corresponding to the variables belonging to the second group and compute the estimated principal components associated to the $r$ largest eigenvalues  as in Section~4. At last,   the total estimator is calibrated on the totals of the variables in $\mathbf{X}_1$ and the first $r$ estimated principal components.
\par

\vspace{1cm}
\setcounter{chapter}{7}
\setcounter{equation}{0} %-1
\noindent {\bf 7. Application to the estimation of the total electricity consumption}

\vspace{0.5cm}

%\section{Application to the estimation of  electricity comsumption} 

\noindent\textbf{Description of the data}

The interest of using principal components calibration is illustrated on data from the Irish Commission for Energy Regulation (CER) Smart Metering Project that was conducted in 2009-2010 (CER, 2011)\footnote{The data are available on request at the address: \texttt{http://www.ucd.ie/issda/data/commissionforenergyregulation/}}.  
In this project, which focuses on energy consumption and energy regulation, about 6000 smart meters have been installed in order to collect every half an hour, over a period of about two years, the electricity consumption of Irish residential and business customers.
 
We evaluate the interest of employing reduction dimension techniques based on PCA by considering a period of 14 consecutive days and  a population of $N=6291$ smart meters (households and companies). Thus, we have for each unit $k$ in the population $(2 \times 7) \times 48 = 672$ measurement instants  and we denote by $y_k(t_j), j=1, \ldots 672$ the data corresponding to unit $k$ where $y_k (t_j)$ is the electricity consumption (in kW) associated to smart meter $k$ at instant $t_j$. We consider here a multipurpose setting and our aim is to estimate the mean electricity consumption of each  day of the second week. For each day $\ell$ of the week, with $\ell \in \{1, \ldots, 7\}$, the outcome variable is $y_{k \ell} = \sum_{j=336 + (\ell-1)\times 48}^{336+ \ell \times 48}y_k  (t_j )$
and our target is the corresponding population total,  $t_{\ell} = \sum_{k \in U} y_{k \ell}$.

The auxiliary information is the load electricity curve of the first week. This means that we have $p=336$ auxiliary variables, which are the consumption electricity levels at each of the $p=336$ half hours of the first week. A sample of 5 auxiliary information curves is drawn in Figure~\ref{fig:excourbes}.
%denoted by  more exactly  $y_k(t_j)$ is the electricity consumption at the instant $t_j, j=1,É,p=336.$  

The condition number of the matrix $N^{-1}\mathbf{X}^T\mathbf{X}$, which is defined as the ratio $\lambda_1/\lambda_{336}$, is equal to 67055.78. This large value means that this matrix is  ill-conditioned and there may exist strong correlations between some of the variables used for calibration. 
Indeed, the first principal component explains about 63\% of the variance of the 336 original variables and about 83\% of the total variability of the data is preserved by projection onto the subspace span by the first ten principal components. %Reducing the dimension should improve the performances of calibration. 

%The cumulative percentage of the total variance is (for the first ten CP)
%\begin{verbatim}
%0.6270056 0.7158621 0.7508449 0.7720227 0.7877237 0.7984318 0.8064813 0.8138370 0.8207524 0.8264292
%\end{verbatim}

\vspace{0.5cm}
\noindent\textbf{Comparison of the estimators}

To make comparisons, we draw $I=1000$ samples of size $n=600$ (the sampling fraction is about $0.095$) 
according to a simple random sampling design without replacement and we estimate the total consumption $t_{\ell}$ over each day $\ell$ of the second week with  the following estimators: 
\begin{itemize}
	\item Horvitz-Thompson estimators, 
	\item Calibration estimators, denoted by  $\hat t_{\ell w}$, 
	that take account of all the $p=336$ auxiliary variables plus the intercept term.
	% $\textbf{X}_j$ (the electricity consumption during the first week) and the intercept and,
	\item Estimators calibrated on the principal components in the population or in the sample plus the constant term,  denoted by $\hat t_{ \ell w}^{\rm{pc}}(r)$, 
	for different values of the dimension $r$.
%	\item the estimated principal components  calibration estimator $\hat t_{\ell w}^{\rm{epc}}(r)$ that takes account of $r$   estimated principal components plus the intercept term. 
\end{itemize}
When performing principal components calibration, the dimension $r$ plays the role of a tuning parameter. We also study the performances of an automatic and simple rule for selecting the dimension $r$ and consider estimators based on principal components  calibration with a data-driven choice of the tuning parameter which consists in  selecting the largest dimension $\hat r$ such that all the estimated principal component weights remain positive. Note that this selection strategy is the analogue of the strategy suggested in Bardsley and Chambers (1984) for choosing the tuning parameter in a ridge regression context.

\begin{figure}
\begin{center}
\includegraphics[width=12cm]{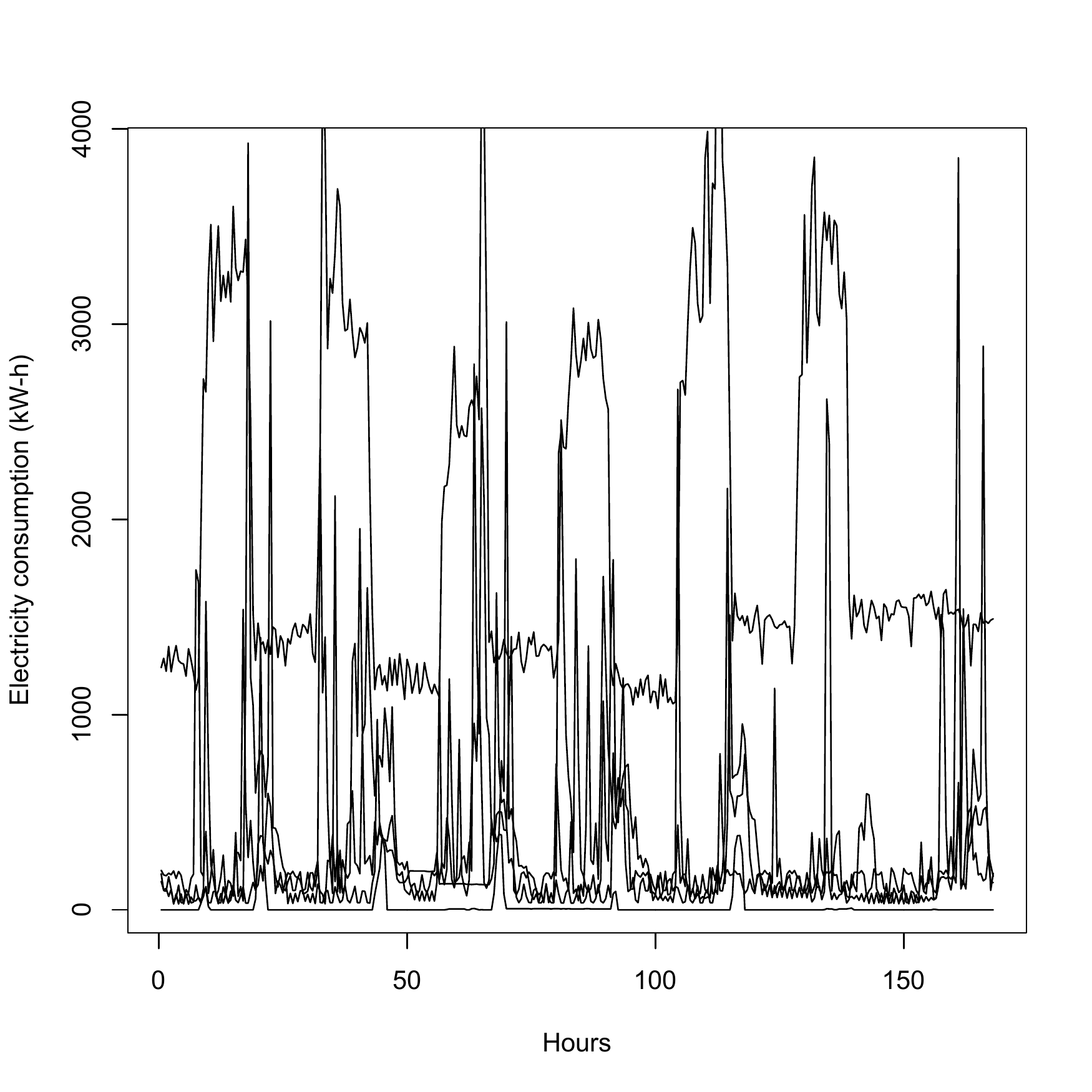}
\caption{A sample of 5 electricity load curves observed every half an hour during the first week.}
\label{fig:excourbes}
\end{center}
\end{figure}

\begin{figure}
\begin{center}
\includegraphics[width=12cm]{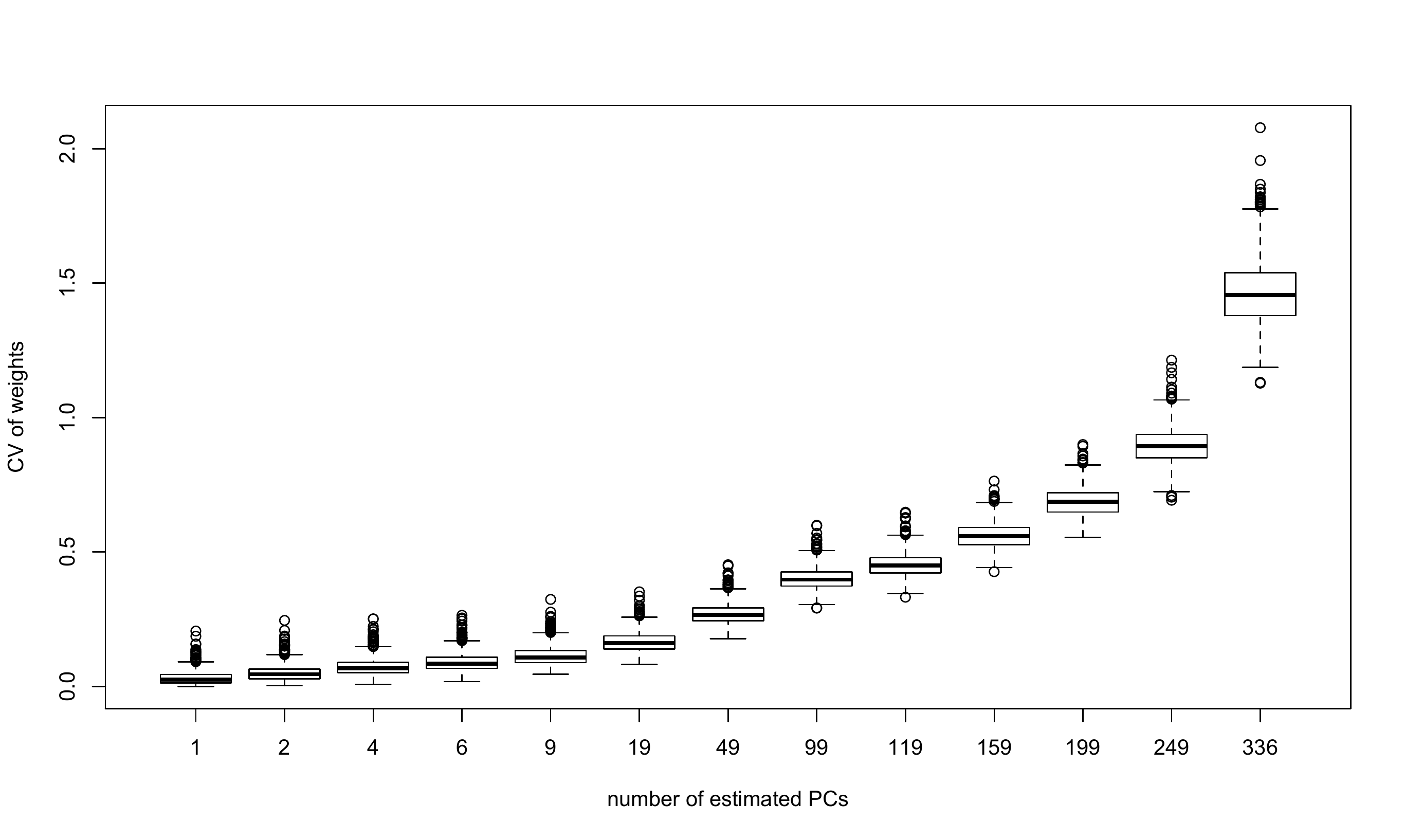}
\caption{Distribution of the coefficient of variation (CV) of the sampling weights for different values of the dimension $r$. The sample size is $n=600$.}
\label{fig:exweight}
\end{center}
\end{figure}

The distribution of the coefficient of variation (CV) of the calibration weights for the $I$=1000 Monte Carlo experiments  is  presented in Figure~\ref{fig:exweight} for different values of the dimension $r$. Recall that these weights do not depend on the variable of interest. It is clearly seen that those calibration weights have larger dispersion and are  more  and more heterogeneous as the  number of principal components used for calibration increases. Calibrating with a large number of correlated auxiliary variables may lead to instable estimations and to a lack of robustness with respect to measurement errors or misspecification in the data bases, with for example the presence of  strata jumpers. Note also that when all the auxiliary variables are used for calibration, around 25 \% of the sampling weights take negatives values which is generally not desirable. The distribution of the proportion of positive sampling weights are given in Figure~\ref{fig:positivew}. It can be seen the proportion of negative weights increases as the number of dimension increases. Note also that it is slightly smaller for the estimated principal components.

\begin{figure}
\begin{center}
\includegraphics[width=16cm]{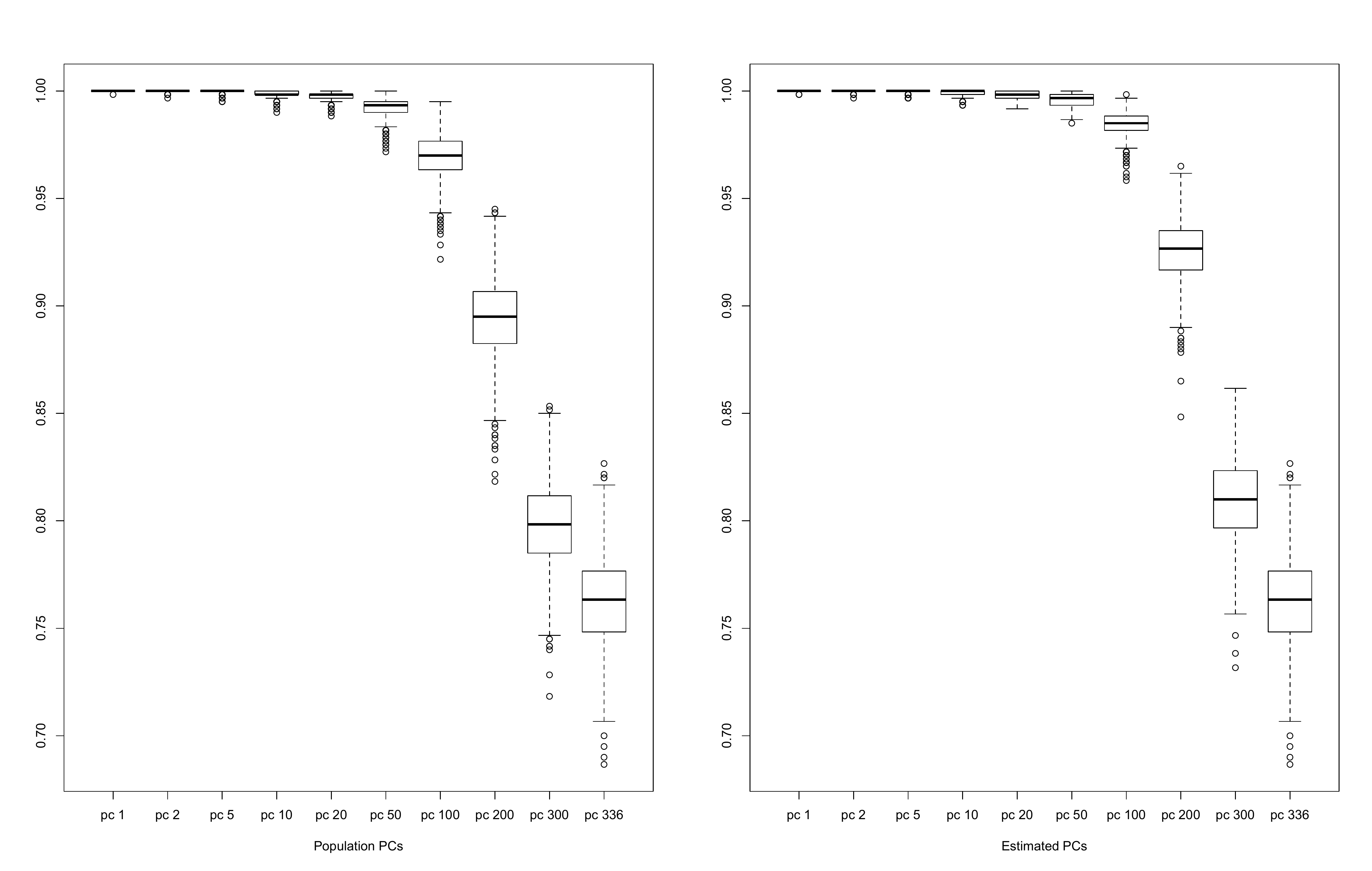}
\caption{Proportion of positive calibrated weights for different values of the dimension~$r$. On the right for calibration on the population principal components and on left for the sample principal components.}
\label{fig:positivew}
\end{center}
\end{figure}

Our benchmarks are the estimators $\hat t_{\ell w}$ which are calibrated on all the $p=336$ auxiliary variables. For each day $\ell$, the performances of an estimator $\widehat{\theta}$ of the total $t_\ell$  are measured by considering the relative mean squared error,
\begin{align}
R_\ell (\widehat{\theta}) &=\frac{\sum_{i=1}^I(\widehat{\theta}^{(i)}-t_\ell )^2}{\sum_{i=1}^I(\widehat{t}_{ \ell w}^{(i)}-t_\ell )^2 }.
\label{def:relMSE}
\end{align}

% Requires the booktabs if the memoir class is not being used
\begin{table}[htbp]
   \centering
   %\topcaption{Table captions are better up top} % requires the topcapt package
   \begin{tabular}{cccccccccc} % Column formatting, @{} suppresses leading/trailing space
      \toprule
  &       & \multicolumn{7}{c}{Days} \\
       \multicolumn{2}{c}{Estimators} & monday & tuesday & wednesday & thursday & friday & saturday & sunday \\
%      \cmidrule(r){2-8} % Partial rule. (r) trims the line a little bit on the right; (l) & (lr) also possible
      \midrule
      Horvitz-Thompson  &  & 14.4 & 13.9 & 11.8 & 10.8 & 12.5 & 6.4 & 5.4 \\ \hline
 $\hat t_{ \ell w}^{\rm{pc}}$ & $r=1$ & 0.65 & 0.62 & 0.50 & 0.47 & 0.64 & 1.17 & 1.57 \\
  $\hat t_{ \ell w}^{\rm{pc}}$ & $r=2$ & 0.64 & 0.62 & 0.50 & 0.47 & 0.57 & 0.80 & 0.63 \\
  $\hat t_{ \ell w}^{\rm{pc}}$ & $r=5$ & 0.52 & 0.47 & 0.40 & 0.50 & 0.51 & 0.53 & 0.52 \\
  $\hat t_{ \ell w}^{\rm{pc}}$ & $r=50$ & 0.50 & 0.50 & 0.43 & 0.44 & 0.54 & 0.48 & 0.48 \\
  $\hat t_{ \ell w}^{\rm{pc}}$ & $r=100$ & 0.57 & 0.60 & 0.59 & 0.51 & 0.58 & 0.60 & 0.64 \\
    $\hat t_{ \ell w}^{\rm{pc}}$ & $r=200$ & 0.60 & 0.64 & 0.58 & 0.66 & 0.69 & 0.68 & 0.63 \\
    $\hat t_{ \ell w}^{\rm{pc}}$ & $r=300$ & 0.82 & 0.85 & 0.83 & 0.86 & 0.84 & 0.85 & 0.87 \\ \hline
 $\hat t_{ \ell w}^{\rm{epc}}$ & $r=1$ & 0.75 & 0.73 & 0.61 & 0.56 & 0.73 & 1.23 & 1.59 \\
  $\hat t_{ \ell w}^{\rm{epc}}$ & $r=2$ & 0.66 & 0.64 & 0.53 & 0.50 & 0.61 & 0.85 & 0.74 \\
  $\hat t_{ \ell w}^{\rm{epc}}$ & $r=5$ & 0.53 & 0.47 & 0.40 & 0.41 & 0.53 & 0.59 & 0.53 \\
  $\hat t_{ \ell w}^{\rm{epc}}$ & $r=50$ & 0.45 & 0.46 & 0.40 & 0.41 & 0.48 & 0.46 & 0.47 \\
  $\hat t_{ \ell w}^{\rm{epc}}$ & $r=100$ & 0.46 & 0.47 & 0.42 & 0.45 & 0.52 & 0.49 & 0.50 \\
    $\hat t_{ \ell w}^{\rm{epc}}$ & $r=200$ & 0.57 & 0.55 & 0.51 & 0.58 & 0.62 & 0.60 & 0.57 \\
    $\hat t_{ \ell w}^{\rm{epc}}$ & $r=300$ & 0.78 & 0.80 & 0.77 & 0.84 & 0.80 & 0.81 & 0.83 \\ \hline
      $\hat t_{ \ell w}^{\rm{pc}}$ &  $\hat r$, $w(\hat r)>0$ & 0.51 & 0.49 & 0.41 & 0.41 & 0.52 & 0.55 & 0.50 \\
        $\hat t_{ \ell w}^{\rm{epc}}$ & $\hat r$, $w(\hat r)>0$ & 0.49 & 0.48 & 0.41 & 0.40 & 0.50 & 0.53 & 0.49 \\ \hline
        Ridge Calibration &  $\hat{\lambda}$ & 0.44 & 0.46 & 0.40 & 0.41 & 0.48 & 0.48 &  0.43 \\ \hline

      \bottomrule
   \end{tabular}
   \caption{Comparison of the mean relative mean squared errors of the different estimators, according to criterion (\ref{def:relMSE}).}
   \label{tab:MSE}
\end{table}

%The HT estimator conducts very bad since $R(\widehat{\hat t_{yd}})=23.3.$\\
Better estimators will correspond to small values of criterion $R_\ell(\widehat{\theta})$.
 The values of this relative error  for several values of $r$, as well as for the estimators obtained with the data driven dimension selection  are given in Table~\ref{tab:MSE}. This relative error has also been computed for the ridge-type  estimators  derived with the sampling weights $w^{\rm{pen}}$ given in Section 2  and a penalty $\hat \lambda$ chosen to be the smallest value of $\lambda$ such that all the resulting weights remain positive.
 
We can first note that the naive Horvitz-Thompson estimator can be greatly improved, for all the days of the week, by considering an over-calibration estimator which takes account of all the (redundant) auxiliary information. Indeed, the mean square error of the HT estimator  is between  five times and fourteen times larger than the MSE of this reference estimator.
We can also remark that reducing the number of effective auxiliary variables through principal components, estimated on the sample or deduced from the population, can still improve estimation compared to calibration on all the variables and permits to divide by two the MSE.  In Figure~\ref{fig:mse} %and Figure~\ref{fig:mseestpc} 
 the MSE is drawn for all the seven considered days and various  dimensions  starting from $r=1$ to $r=336$. Note that the largest dimension corresponds to the estimators calibrated on all the original variables. 
A remarkable feature is the stability of the  principal components calibration techniques with respect to the choice of the dimension $r$. Indeed, in this application, choosing between 5 and 100 principal components permits to divide by two, for all the outcome variables, the MSE compared to the calibration estimator based on the whole auxiliary information. 

% The distribution of the number of selected principal components is drawn in Figure~\ref{fig:dim}. 
 We noted that the mean number of selected principal components with the data driven selection rule is equal to 17.3  for the population principal components and 21.3 for the sample principal components, explaining in each case about 85\%  of the variance of the original variables.
As expected, the variability of the number of selected components is slightly larger when considering calibration on the estimated principal components (interquartile range of 26 versus 17 for the population principal components).

% \begin{figure}
%\begin{center}
%\includegraphics[width=14cm]{CalibErrorPoPPC.pdf}
%\includegraphics[width=9cm]{boxplotdim.pdf}
%\caption{Distribution of the number of selected principal components used for calibration.}
%\label{fig:dim}
%\end{center}
%\end{figure}

As seen in Table~\ref{tab:MSE}, the performances of the resulting estimators are good and comparable to the estimators based on ridge calibration with a selection rule for $\lambda$ based on the same principle. The advantage of the principal components is that  it permits to divide by more than 15 the final number of effective variables used for calibration and it can directly be used  in classical survey sampling softwares. 
 
\begin{figure}
\begin{center}
\includegraphics[width=15cm]{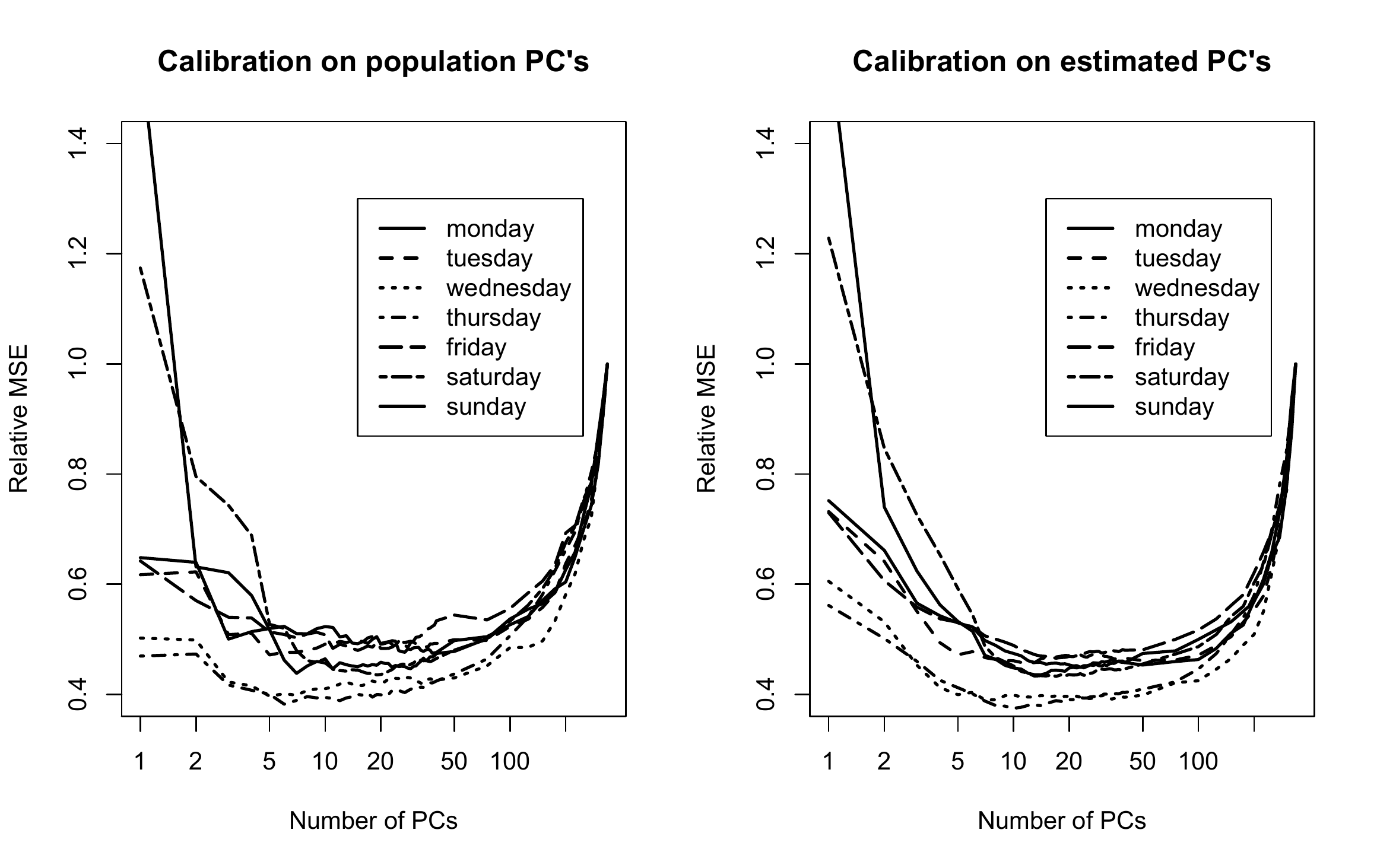}
\caption{Relative MSE, defined in (\ref{def:relMSE}), for different values of the dimension $r$ and the different days of the week and a sample size  of $n=600$ for calibration estimators based on the population (left) and the sample (right) principal components. The horizontal axis is at a log scale.}
\label{fig:mse}
\end{center}
\end{figure}

Finally, we have also evaluated the calibration error. It is  the difference between the total of the original auxiliary variables and their "estimation" on the samples obtained with the weights  $w_k^{\rm{pc}}(r)$ defined in (\ref{pcestimator}) and the weights $w_k^{\rm{epc}}(r)$ defined in (\ref{hatpcestimator}).  The distribution the estimation squared calibration error $\| \sum_{k \in s} w_k \mathbf{x}_k - t_{\mathbf{x}}\|^2$  is drawn in Figure~\ref{fig:txpop} (resp. Figure~\ref{fig:txsample}) for calibration on  principal components at the population level (resp. for the estimated principal components). We have also plotted the distribution (first boxplot) of the squared error for the weight obtained with the data driven choice of the dimension $r$. For both approaches, the distributions of the errors are very similar. The errors 	are high and highly variable when the number of principal components is small (a mean value of about 1300 for $r=1$) and then they decrease rapidly  (a mean value of 720 for $r=5$ and 600 for $r=10$). For larger values of $r$, the decrease is slower. When the dimension $r$ is not chosen in advance, the mean value is about at the level of the error when is equal to $10$ but with a variability  that is much larger.

 \begin{figure}
\begin{center}
\includegraphics[width=15cm]{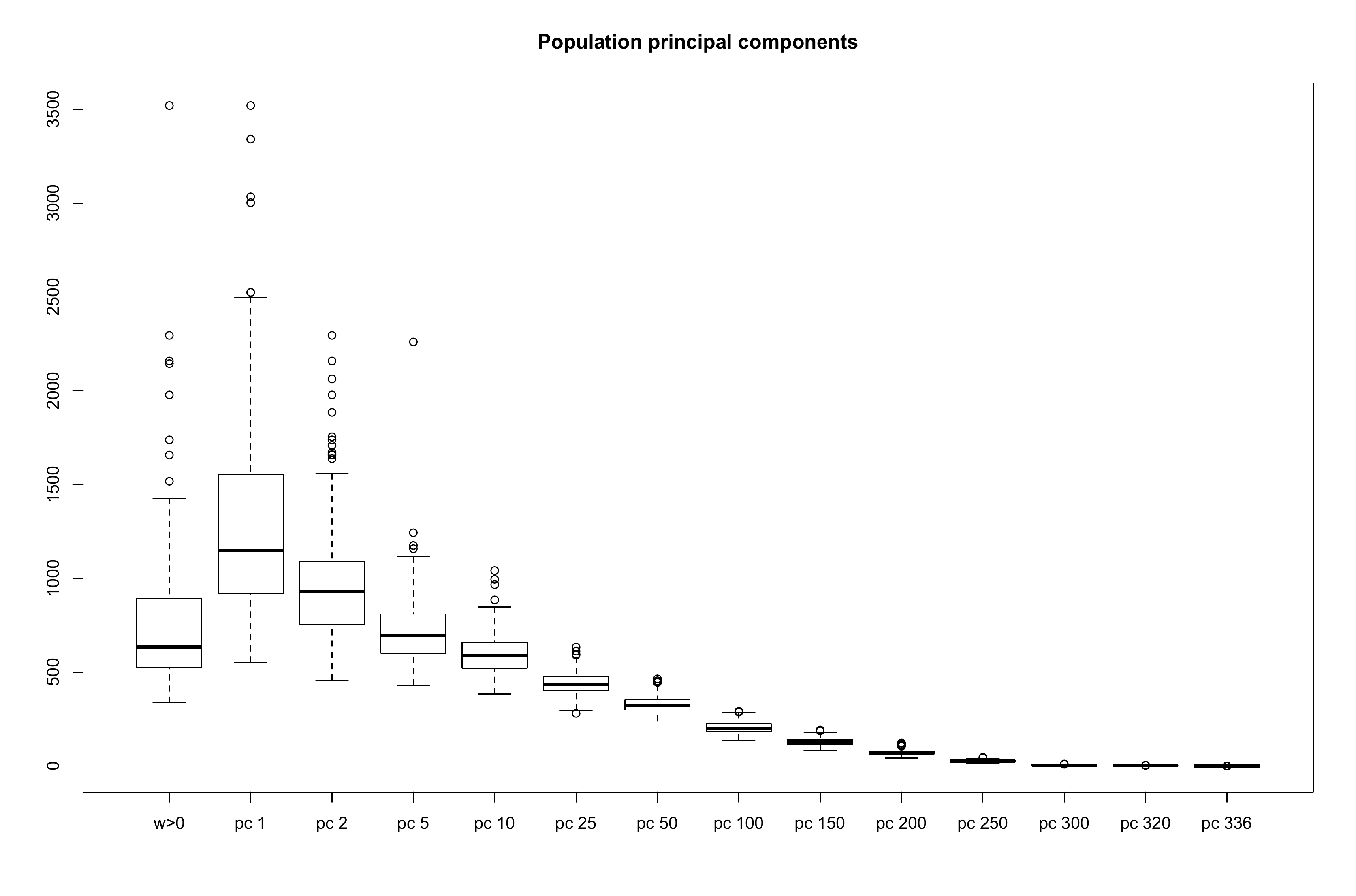}
\caption{Calibration error for the original variables, in terms of MSE, for different number of principal components estimated at the population level. The first boxplot ($w>0$) corresponds to the data-driven choice of the number of components.}
\label{fig:txpop}
\end{center}
\end{figure}

 \begin{figure}
\begin{center}
\includegraphics[width=15cm]{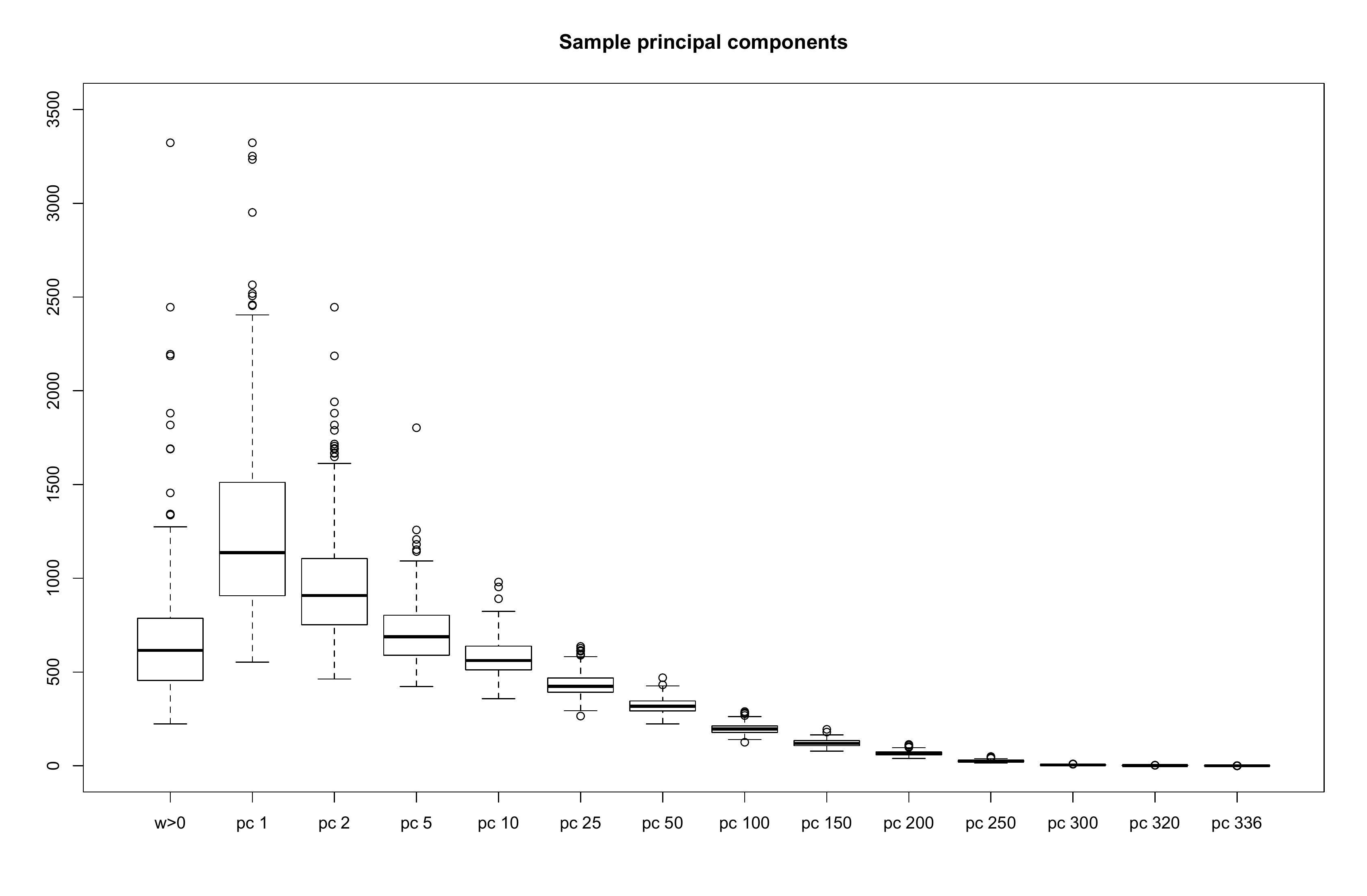}
\caption{Calibration error for the original variables, in terms of MSE, for different number of principal components estimated in the sample. The first boxplot ($w>0$) corresponds to the data-driven choice of the number of components.}
\label{fig:txsample}
\end{center}
\end{figure}
%\begin{figure}
%\begin{center}
%\includegraphics[width=14cm]{CalibErrorEstPC.pdf}
%\caption{Evolution of the relative MSE, defined in (\ref{def:relMSE}), for different values of the dimension $r$ and the different days of the week and a sample size  of $n=600$ for calibration estimators based on the sample principal components. The horizontal axis is at a log scale.}
%\label{fig:mseestpc}
%\end{center}
%\end{figure}
%\section{Discussion and concluding remarks}
\clearpage 
\vspace{1cm}
\setcounter{chapter}{8}
\setcounter{equation}{0} %-1

\noindent {\bf 8. Discussion and concluding remarks}

\vspace{0.5cm}

A simple dimension reduction technique based on principal components calibration has been studied in this article. It provides an effective technique for approximate calibration, when the number of auxiliary variables is large, that can improve significantly the estimation compared to calibration on the whole set of initial auxiliary variables. Furthermore this simple technique can also be modified so that calibration can be exact for a set of a few important auxiliary variables. 
A very simple rule which consists in choosing the largest number of principal components such that all  the resulting sampling weights remain positive allows to construct estimators with very good performances in terms of MSE. Furthermore, in our case study, this automatic selection technique allows, in average, to divide by 15  the number of effective calibration variables and to reduce by a half the mean squared error of estimation compared to calibration over all the auxiliary variables. 

%We have also  noted in the illustration on electricity consumption   that a "bad" choice of the number of principal components which are used as calibration variables may not have very important consequences. Indeed, the issue of choosing adequately the dimension $r$ for calibration when the aim is to estimate totals  is not as dramatic as in principal components regression when the aim is to estimate the "true" regression coefficient. However, finding  automatic data-driven procedures that could help in choosing among the best values for the dimension $r$ is of  interest.  Aggregated criterions based on cross-validation (see Opsomer and Miller, 2005) or penalized cross-validation for principal components regression  could certainly be adapted to our multipurpose and finite population context. Note also that having robustness issues in mind, considering a criterion that could also take  account  of the stability of the calibration weights would certainly be of interest. This issue clearly deserves further investigation and is beyond the scope of the present work.

Some asymptotic justifications of this dimension reduction technique are given with a number $p_N$ of auxiliary variables as well as a number $r_N$ of principal components used for calibration that are allowed to grow to infinity as the population size $N$ increases. Nevertheless, our conditions on the asymptotic behavior of $r_N$ appear to be rather restrictive and could probably be relaxed (see for example the results presented in Vershynin, 2012 on the estimation of covariance matrices for independent observations in a infinite population). However, this would require to have at hand exponential inequalities for Horvitz-Thompson estimators in order to control in  very accurately their deviations around their target. This difficult issue, in a finite population context, is clearly beyond the scope of this paper but certainly deserves further investigation.   

 %It would be interesting to examine what happens when the number $p$ of auxiliary variables is  allowed to tend to infinity when the population and the sample  size grow. Different situations about the asymptotic behavior of the smallest eigenvalue of matrix $(Np)^{-1} \mathbf{X}^T \mathbf{X}$ might   be distinguished. According to the fact that it tends to zero (highly correlated data) or not (nearly independent data), different conclusions on the rates of convergence and how the number of principal components should be chosen may be drawn. This difficult issue is related to functional data analysis and inverse problems techniques and is beyond the scope of the paper. 

 Note finally that borrowing ideas from Marx and Smith (1990) and Wu and Sitter (2001), it would be not too difficult to extend the principal component calibration approach in order to deal with non linear model-calibration.
 
\medskip

\noindent \textbf{Acknowledgements}. The authors thank an associate editor  and the anonymous referees  for comments and suggestions, and more particularly for suggesting the data-driven  rule for selecting the number of components.

%\subsection*{Selecting the number of principal components}
%%%%%%%%%%%%%%%%%%%%%%%%%%%%%%%%%%%%%
%%%%%%%%%%%%%%%%%%%%%%%%%%%%%%%%%%%%%

%The dimension $r$ plays the role of a tuning parameter and the user has to find  a good compromise between exact calibration, $r=p$, that may lead to high and unstable weights and  $r=0,$ which means that no auxiliary information has been used.
%Jolliffe (2002) provides criteria for choosing the number of principal components. The screeplot

%We did not suppose that $p$ could tend to infinity. 
%Different cases. 
%\begin{itemize}
%\item Independent and nearly independent case : the smallest eigenvalue $\lambda_p$ of matrix $\frac{1}{Np} \mathbf{X}^T \mathbf{X}$ satisfies $\lambda_p \geq \lambda_{\min} >0$ for all $p > p_0$
%\item Correlated case  : $\lambda_p \to 0$. For a detailed study of the correlated case (once can think of functional data discretized at equidistant time instants), we must know a bit more on how fast $\lambda_p$ tends to zero.
 % \end{itemize}
%%%%%%%%%%%%%%%%%%%%%%%%%%%%%%%%%%%%%%%%%%%%%%%%%%%%%%%%%%%
%%%%%%%%%%%%%%%%%%%%%%%%%%%%%%%%%%%%%%%%%%%%%%%%%%%%%%%%%%%
%\Appendix
%%%%%%%%%%%%%%%%%%%%%%%%%%%%%%%%%%%%%%%%%%%%%%%%%%%%%%%%%%%
%%%%%%%%%%%%%%%%%%%%%%%%%%%%%%%%%%%%%%%%%%%%%%%%%%%%%%%%%%%
\section*{References}
%%%%%%%%%%%%%%%%%%%%%%%%%%%%%%%%%%%%%%%%%%%%%%%%%%%%%%%%%%%Skinner and Silva, 1997; 
%%%%%%%%%%%%%%%%%%%%%%%%%%%%%%%%%%%%%%%%%%%%%%%%%%%%%%%%%%%
\begin{description}

\item Bardsley, P. and Chambers, R. (1984). Multipurpose estimation from unbalanced samples. \textit{Applied Statistics}, \textbf{33}, 290-299.

\item Beaumont, J.-F. and Bocci, C. (2008). Another look at ridge calibration.
{\em Metron-International Journal of Statistics, LXVI}: 5-20.

\item Bosq, D. (2000). {\em  Linear processes in function spaces}. Lecture Notes in Statistics, Vol. 149. Springer-Verlag.

\item  Breidt, J.F. and Opsomer, J. D.(2000). Local polynomial regression estimators in survey sampling. {\em Ann.  Statist.}, {\bf 28}, 1023-1053.

\item Breidt, J.F. and Chauvet, G. (2012). Penalized balanced sampling. \textit{Biometrika}, \textbf{99}, 945-958.

\item Cardot, H., Chaouch, M., Goga, C. and  Labru\`ere, C. (2010). Properties of design-based functional principal components analysis. \textit{Journal of Statistical and Planning Inference}, \textbf{140}, 75-91.

\item Cassel, C., S\"arndal, C.-E. and Wretman, J. (1976). Some results on generalized difference estimation and generalized regression estimation for finite populations. \textit{Biometrika}, \textbf{63}, 615-620.  
 
\item Chambers, R.L. (1996). Robust case-weighting for multipurpose establishment surveys. {\em Journal of Official Statistics,} \textbf{12}, 3-32.

\item Chambers, R.,  Skinner, C. and Wang, S. (1999). Intelligent calibration. \textit{Bulletin of the International Statistical Institute}, \textbf{58(2)}, 321-324.

\item Chen, J., Sitter, R.R. and Wu, C. (2002). Using empirical likelihood methods to obtain range restricted weights in regression estimators for surveys. \textit{Biometrika}, \textbf{89}, 230-237.

\item Clark, R. G. and Chambers, R.L. (2008). Adaptive calibration for prediction of finite population totals. \textit{Survey Methodology}, \textbf{34}, 163-172.

\item Deville, J.C. and S\"arndal, C.E. (1992). Calibration estimators in survey sampling.  {\em J. Amer. Statist. Assoc.}, \textbf{87}, 376-382. 

\item Frank, I.E. and Friedman, J.H. (1993). A statistical view of some chemometrics regression tools. {\it Technometrics}, {\bf 35}, 109-148.

\item Goga, C., Shehzad, M.A., and Vanheuverzwyn, A. (2011). Principal component regression with survey data. application on
the french media audience. {\em Proceedings of the 58th World Statistics Congress of the International Statistical Institute}, Dublin, Ireland, 3847-3852.

\item Goga, C. and Shehzad, M.A. (2014). A note on partially penalized calibration. \textit{Pakistan Journal of Statistics}, \textbf{30}(4), 429-438. 
 
\item Guggemos, F. and Till\'e, Y. (2010). Penalized calibration in survey sampling: Design-based estimation assisted by mixed models.
{\em Journal of Statistical Planning and Inference}, \textbf{140}, 3199-3212.

\item Gunst, R.F. and Mason, R.L. (1977). Biased estimation in regression: an evaluation using mean squared error.
 {\em J. Amer. Statist. Assoc.}, \textbf{72}, 616-628.

\item Hoerl, E. and Kennard, R. W. (1970). Ridge Regression: Biased Estimation for Nonorthogonal Problems.  \textit{Technometrics}, 12, 55-67.

\item Isaki, C. and Fuller, W. (1982). Survey design under the regression superpopulation model.
 {\em J. Amer. Statist. Assoc.}, {\bf 77}, 49-61.

\item Jolliffe, I.T. (2002). \textit{Principal Components Analysis}. Second Edition, Springer- Verlag. New York.

\item Rao, J.N.K. and Singh, A.C. (1997). A ridge-shrinkage method for range-restricted weight calibration in survey sampling.
{\em Proceedings of the Section on Survey Research Methods, American Statistical Association}, 57-65.

\item Ren, R. (2000). {\em Utilisation de l'information auxiliaire par calage sur fonction de r\'epartition}. PhD, Universit\'e Paris Dauphine, France.

\item S\"arndal, C.E., Swensson, B., and Wretman J. (1992). \textit{Model Assisted Survey Sampling}. Springer-Verlag, New York Inc.

\item S\"arndal, C.E. (2007). The calibration approach in survey theory and practice. {\em Survey Methodology}, {\bf 33}, 99-119.

\item Shehzad, M.-A. (2012). {\em Penalization and auxiliary information reduction methods in surveys.} Doctoral dissertation, Universit\'e de Bourgogne, France. \\ Available at {\small \texttt{http://tel.archives-ouvertes.fr/tel-00812880/.}}

\item Silva, P.L.N. and Skinner, C. (1997). Variable selection for regression estimation in finite populations. {\em Survey Methodology}, \textbf{23}, 23-32.

\item  Singh, A. C. and Mohl, C. (1996). Understanding calibration in survey sampling. {\em Survey Methodology}, \textbf{22}, 107-115.

\item  Swold, S., Sj\"ostr\"om, M. and Eriksson, L. (2001). PLS-regression : a basic tool of chemometrics. {\em Chemometrics and Intelligent Laboratory},
{\bf 58}, 109-130.

\item Tibshirani, R. (1996). Regression shrinkage and selection via the lasso. {\em J. Roy. Statist. Soc. Ser. B}, \textbf{58}, 267-288.

\item Th\'eberge, A. (2000). Calibration and restricted weights. \textit{Survey Methodology}, \textbf{26}, 99-107.

\item Vershynin, R. (2012). How close is the sample covariance matrix to the actual covariance matrix? 
{\em J. Theoret. Probab.} {\bf 25}, 655-686. 

\item Wu, C. and Sitter, R. R. (2001). A Model-Calibration Approach to Using Complete Auxiliary Information from Survey Data. {\em J. Amer. Statist. Assoc.}, {\bf 96}, 185-193.

\item Zou, H. and Hastie, T. (2005). Regularization and variable selection via the Elastic Net. {\em J. Roy. Statist. Soc. Ser. B}, \textbf{67}, 301-320.

\end{description}

\vskip .65cm
\noindent Cardot, H.\\
IMB, UMR CNRS 5584, Universit\'e de Bourgogne, 9 avenue Alain Savary, Dijon, France 
\vskip 2pt
\noindent herve.cardot@u-bourgogne.fr\\
\vskip 2pt
\noindent Goga, C.\\
IMB, UMR CNRS 5584, Universit\'e de Bourgogne, 9 avenue Alain Savary, Dijon, France 
\vskip 2pt
\noindent camelia.goga@u-bourgogne.fr\\
\vskip 2pt
\noindent Shehzad, M.-A.\\
Bahauddin Zakariya University, Bosan Road Multan, Pakistan.
\vskip 2pt
\noindent mdjan708@gmail.com
\vskip .3cm

\clearpage

\section*{Appendix : Proofs}

Throughout the proofs, we use the letter C to denote a generic constant whose value may vary from place to place. This constant does not depend on N.  For sake of clarity, subscript $_N$ has been suppressed when there were no ambiguity. 
%\end{appendix}

For a vector $\mathbf{v}$, we denote by $\| \mathbf{v}\|$ its Euclidean norm. The spectral norm of a matrix $\mathbf{A}$ is denoted by $\| \mathbf{A} \| = \sup_{ \mathbf{v}} \| \mathbf{Av} \| / \|\mathbf{v} \|$. We often use the following  well known inequality,  $\| \mathbf{A} \|^2 \leq \mbox{tr} \left( \mathbf{A}^T \mathbf{A} \right)$ as well as the equality $\| \mathbf{A}^T \mathbf{A} \| = \| \mathbf{A}\mathbf{A}^T\|$.

\subsection*{Proof of Proposition 1} 
We may write 
\begin{align}
 \hat{t}_{yw}^{\rm{pc}} (r) - t_y  &= \left( \tilde t_{y,\mathbf{x}}^{\rm{diff}}(r) - t_y \right) + \left(\hat{t}_{\mathbf{z}_rd}-t_{\mathbf{z}_r}\right)^T \left(\hat{\boldsymbol{\gamma}}_{\mathbf{z}}(r)-\tilde{\boldsymbol{\gamma}}_{\mathbf{z}}(r) \right),
\label{dec:tzdiff}
\end{align}
where  
\begin{align}
\tilde t_{y,\mathbf{x}}^{\rm{diff}}(r) &=\hat{t}_{yd}-\left(\hat{t}_{\mathbf{z}_rd}-t_{\mathbf{z}_r}\right)^T \tilde{\boldsymbol{\gamma}}_{\mathbf{z}}(r) .
\label{def:tydiff}
\end{align}

We get by linearity of the Horvitz-Thompson estimators, that
\begin{align}
\left(\hat{t}_{\mathbf{z}_rd}-t_{\mathbf{z}_r}\right)^T \tilde{\boldsymbol{\gamma}}_{\mathbf{z}}(r) &= \widehat{t}_{ \mathbf{z}_r^T\tilde{\boldsymbol{\gamma}}_{\mathbf{z}}(r)}  - t_{ \mathbf{z}_r^T\tilde{\boldsymbol{\gamma}}_{\mathbf{z}}(r)}
\label{gammalinearity}
\end{align}
By construction, the new real variable $\mathbf{z}_r^T\tilde{\boldsymbol{\gamma}}_{\mathbf{z}}(r)$  is a projection and we have that 
\begin{align*}
\sum_{k \in U_N} \left(\mathbf{z}_{kr}^T\tilde{\boldsymbol{\gamma}}_{\mathbf{z}}(r)\right)^2 & \leq \sum_{k \in U_N} y_k^2.
\end{align*}
Consequently, we get with~(\ref{gammalinearity}) and with classical properties of Horvitz-Thompson estimators (and because assumptions (A1) and (A2)  hold) that  
\begin{align}
\frac{1}{N} \left(\hat{t}_{\mathbf{z}_rd}-t_{\mathbf{z}_r}\right)^T \tilde{\boldsymbol{\gamma}}_{\mathbf{z}}(r) & = O_p \left(\frac{1}{\sqrt{n}} \right).
\label{bound:tz2}
\end{align}

%We have that $\tilde t_{y,\mathbf{x}}^{\rm{diff}}(r_N) - t_y = (\hat{t}_{yd} - t_y) -\left(\hat{t}_{\mathbf{x}d}-t_{\mathbf{x}}\right)^T\tilde{\boldsymbol{\beta}}^{\rm{pc}}_{\mathbf{x}}(r_N)$. With assumptions (A1) to (A3), we have that  $(\hat{t}_{yd} - t_y) =O_p(n^{-1/2})$.
%Let   $\mathbf{G}_{r}=(\mathbf{v}_1, \ldots, \mathbf{v}_{r})$ be the $N\times r$ matrix whose columns are the  $r$ orthonormal eigenvectors associated to the $r$ largest eigenvalues of $\left(\mathbf{X}^T\mathbf{X}\right)/N$. 
Recall that $(\mathbf{Z}_1, \ldots, \mathbf{Z}_r)=\mathbf{X}\mathbf{G}_r$, and 
\[
N^{-1}\sum_{k\in U}\mathbf{z}_{kr}\mathbf{z}^T_{kr} =  \mathbf{G}^T_r\left(N^{-1}\mathbf{X}^T\mathbf{X}\right)\mathbf{G}_r =  \mbox{diag}(\lambda_j)_{j=1}^r:=\boldsymbol{\Lambda}_r.
\]
By definition of the principal components and assumption (A5) we have that $\sum_{k \in U_N} \| \mathbf{z}_{kr} \|^2 = \left( \sum_{j=1}^{r} \lambda_j \right) N \leq C N r $. 
%where $P_{r}$ is the orthogonal projection onto the vector space spanned by $\mathbf{v}_1, \ldots, \mathbf{v}_{r}$
%$N^{-1}(\hat t_{yw}^{\rm{pc}}(r)-\tilde t_{y,\mathbf{x}}^{\rm{diff}}(r)) =N^{-1}\left(\hat{t}_{\mathbf{x}d}-t_{\mathbf{x}}\right)^T(\hat{\boldsymbol{\beta}}^{\rm{pc}}_{\mathbf{x}}(r)-\tilde{\boldsymbol{\beta}}^{\rm{pc}}_{\mathbf{x}}(r))$.
%Noting that 
%\begin{align*}
%\tilde t_{y,\mathbf{x}}^{\rm{diff}}(r)) &= %\end{align*}
Following the same lines as in Breidt and Opsomer (2000), we obtain with assumptions (A1)-(A5) that 
\begin{align}
\frac{1}{N} \left(\hat{t}_{\mathbf{z}_rd}-t_{\mathbf{z}_r}\right) &=O_p\left(\sqrt{\frac{r}{n}} \right).
\label{bound:tz}
\end{align}

It remains to bound $\hat{\boldsymbol{\gamma}}_{\mathbf{z}_r}-\tilde{\boldsymbol{\gamma}}_{\mathbf{z}_r}$. For that, we first bound the estimation error of $\boldsymbol{\Lambda}_r$. %Denote by $\widehat{\boldsymbol{\Lambda}}_r = N^{-1} \sum_{k \in s} d_k \mathbf{z}_{kr}\mathbf{z}_{kr}^T$. 
Considering the spectral norm for square matrices, we have for some constant $C$, with classical algebra (see \textit{e.g.} Cardot \textit{et al.} 2010, Proposition 1), 
\begin{align}
E_p \left\| \boldsymbol{\Lambda}_r -  \frac{1}{N} \sum_{k \in s} d_k \mathbf{z}_{kr}\mathbf{z}_{kr}^T \right\|^2 & \leq \frac{C}{N^2} \sum_{k \in U_N} \left\| \mathbf{z}_{kr} \right\|^4.
\label{bound:gammaz}
\end{align}
Expanding now each $\mathbf{z}_{kr}$ in the eigenbasis $\mathbf{v}_1, \ldots, \mathbf{v}_r$, we have 
$\| \mathbf{z}_{kr} \|^2 = \sum_{j=1}^r \langle x_k, v_j \rangle^2$ and thus 
\begin{align}
\frac{1}{N} \sum_{k \in U_N} \left\| \mathbf{z}_{kr} \right\|^4 &\leq \frac{r}{N}  \sum_{k \in U_N} \sum_{j=1}^r \langle x_k, v_j \rangle^4 \nonumber \\
 & \leq  r \sum_{j=1}^r \left[ \frac{1}{N}  \sum_{k \in U_N} \langle x_k, v_j \rangle^4 \right] \nonumber \\
 & \leq C_4 r^2,
 \label{bound:z4}
\end{align}
thanks to assumption (A6). We deduce from (\ref{bound:gammaz}) and (\ref{bound:z4}) that
\begin{align}
\left\| \boldsymbol{\Lambda}_r - \frac{1}{N} \sum_{k \in s} d_k \mathbf{z}_{kr}\mathbf{z}_{kr}^T \right\| &= O_p \left( \frac{r}{\sqrt{n}} \right). 
\label{err:gamma}
\end{align}
Note that as in Cardot \textit{et al.} (2010), we can deduce from previous upper bounds that 
\begin{align}
\left\| \left(\frac{1}{N} \sum_{k \in s} d_k \mathbf{z}_{kr}\mathbf{z}_{kr}^T\right)^{-1} \right\| &= \frac{1}{\lambda_r} + O_p \left(  \frac{r}{\sqrt{n}} \right),
\label{maj:hatinvlambda}
\end{align}
and 
\begin{align}
\left\| \boldsymbol{\Lambda}_r^{-1}  - \left(\frac{1}{N} \sum_{k \in s} d_k \mathbf{z}_{kr}\mathbf{z}_{kr}^T\right)^{-1} \right\| & \leq \frac{1}{\lambda_r} \left\| \left(\frac{1}{N} \sum_{k \in s} d_k \mathbf{z}_{kr}\mathbf{z}_{kr}^T\right)^{-1}\right\| \left\| \boldsymbol{\Lambda}_r  - \left(\frac{1}{N} \sum_{k \in s} d_k \mathbf{z}_{kr}\mathbf{z}_{kr}^T\right) \right\| \nonumber \\
&= O_p \left( \frac{r}{\sqrt{n}} \right).
\label{maj:difflambda}
\end{align}

An application of Cauchy-Schwarz inequality as well as the bound obtained in~(\ref{bound:z4}), gives with assumptions (A1)-(A6) that there is some constant $C$ such that 
\begin{align}
\frac{1}{N^2} E_p \left\| \sum_{k \in U_N} y_k \mathbf{z}_{kr} -\sum_{k \in s} d_k y_k \mathbf{z}_{kr} \right\|^2 & \leq \frac{C}{N^2} \sum_{k \in U_N} y_k^2 \left\| \mathbf{z}_{kr} \right\|^2 \nonumber \\
& \leq \frac{C}{N^2} \left(\sum_{k \in U_N} y_k^4\right)^{1/2} \left(\sum_{k \in U_N} \left\| \mathbf{z}_{kr} \right\|^4 \right)^{1/2} \nonumber \\
& \leq C \frac{r}{N}.
\label{bound:deltaz}
\end{align} 

%Writing now $\tilde{\boldsymbol{\gamma}}_{\mathbf{z}}(r) = \left(\boldsymbol{\Lambda}_r\right)^{-1} \left(\frac{1}{N}  \sum_{k \in U_N}  \mathbf{z}_{kr} y_k \right)$,

Note finally, that we have for some constant $C$
\begin{align}
\left\| \frac{1}{N} \sum_{k \in U_N} \mathbf{z}_{kr} y_k \right\|^2 &=  \frac{1}{N^2} \sum_{k, \ell \in U_N} y_k y_\ell  \mathbf{z}_{kr}^T \mathbf{z}_{\ell r} \nonumber \\
 & \leq  \lambda_1 \frac{1}{N} \sum_{k \in U_N} y_k^2  \nonumber \\
 & \leq C,
%\label{equ:gammatildebounded} 
\end{align}
because the largest eigenvalue of the non negative $N \times N$ matrix ${N}^{-1} \mathbf{XG}_r \left(\mathbf{XG}_r\right)^T$ is equal to $\lambda_1$ and ${N}^{-1} \sum_{k \in U_N} y_k^2$ is supposed to be bounded.

Consequently, we get with previous upper bounds,
\begin{align}
\left\| \hat{\boldsymbol{\gamma}}_{\mathbf{z}}(r)-\tilde{\boldsymbol{\gamma}}_{\mathbf{z}}(r) \right\| & \leq  \left\| \boldsymbol{\Lambda}_r^{-1} - \left(\frac{1}{N} \sum_{k \in s} d_k \mathbf{z}_{kr}\mathbf{z}_{kr}^T\right)^{-1} \right\| \left\| \frac{1}{N}  \sum_{k \in U_N} \mathbf{z}_{kr} y_k \right\|  \nonumber \\
  &+ \left\| \frac{1}{N}  \sum_{k \in U_N} \mathbf{z}_{kr} y_k  - \frac{1}{N}  \sum_{k \in s} d_k \mathbf{z}_{kr} y_k \right\| \left\| \left(\frac{1}{N} \sum_{k \in s} d_k \mathbf{z}_{kr}\mathbf{z}_{kr}^T\right)^{-1} \right\| \nonumber \\
&= O_p \left( \frac{r}{\sqrt{n}} \right)  
\label{bound:gammazest}
\end{align}
and with (\ref{bound:tz}),
\begin{align}
\frac{1}{N} \left(\hat{t}_{\mathbf{z}_rd}-t_{\mathbf{z}_r}\right)^T \left(\hat{\boldsymbol{\gamma}}_{\mathbf{z}}(r)-\tilde{\boldsymbol{\gamma}}_{\mathbf{z}}(r) \right) &= O_p \left( \frac{r^{3/2}}{n} \right).
\end{align}
Finally, using again decomposition (\ref{dec:tzdiff}), we get with previous bounds
\begin{align}
\frac{1}{N} \left( \hat{t}_{yw}^{\rm{pc}} (r) - t_y \right) &= \frac{1}{N} \left( \tilde t_{y,\mathbf{x}}^{\rm{diff}}(r) - t_y \right) + O_p \left( \frac{r^{3/2}}{n} \right) . 
\end{align}
% See also the proof of Proposition~\ref{result_pc_estim} below.
\hfill $\square$

\subsection*{Proof  of Proposition 2}

The proof follows the same lines as the proof of Proposition 1. We first write 
\begin{align}
\frac{1}{N} \left(\hat t_{yw}^{\rm{epc}}(r)-t_y \right) &=  \frac{1}{N}  \left(\tilde t_{y,\mathbf{x}}^{\rm{diff}}(r)-t_y\right) + \frac{1}{N} \left(\hat{t}_{\mathbf{x}d} -t_{\mathbf{x}} \right)^T \left( \tilde{\boldsymbol{\beta}}^{\rm{pc}}_{\mathbf{x}}(r) - 
\hat{\boldsymbol{\beta}}^{\rm{epc}}_{\mathbf{x}}(r) \right), 
\label{dec:tzhatdiff}
\end{align}
and note that, as in Proposition 1, we have that $N^{-1}  \left(\tilde t_{y,\mathbf{x}}^{\rm{diff}}(r)-t_y\right) = O_p(1/\sqrt{n})$. 

We now look for an upper bound on the second term at the right-hand side of equality (\ref{dec:tzhatdiff}). %where $\hat t_{y,\mathbf{x}}^{\rm{diff}}(r)$ is defined as follows
%\begin{align}
%\hat t_{y,\mathbf{x}}^{\rm{diff}}(r) & = \hat{t}_{yd}-\left(\hat{t}_{\hat{\mathbf{z}}_rd}-t_{\hat{\mathbf{z}}_r}\right)^T \tilde{\boldsymbol{\gamma}}_{\mathbf{z}}(r) .
%\label{def:tyhatdiff}
%\end{align}
%&= N^{-1}\left(\tilde t_{y,\mathbf{x}}^{\rm{diff}}(r)-t_y \right)+o_p(n^{-1/2})$ and the results follows from the above proposition.
 It can be shown easily that $N^{-1}\left( \hat{t}_{\mathbf{x}d} -t_{\mathbf{x}}  \right) = O_p( \sqrt{p/n})$.

We can also write 
\begin{align}
\tilde{\boldsymbol{\beta}}^{\rm{pc}}_{\mathbf{x}}(r) -  \hat{\boldsymbol{\beta}}^{\rm{epc}}_{\mathbf{x}}(r) &= \mathbf{G}_r \left(\tilde{\boldsymbol{\gamma}}_{\mathbf{z}}(r) - \hat{\boldsymbol{\gamma}}_{\hat{\mathbf{z}}}(r) \right) + \left( \mathbf{G}_r  - \hat{\mathbf{G}}_r \right) \hat{\boldsymbol{\gamma}}_{\hat{\mathbf{z}}}(r) 
\end{align}
and bound each term at the right-hand side of the equality.

We  denote by $\hat{\mathbf{G}}_r=(\hat{\mathbf{v}}_1, \ldots, \hat{\mathbf{v}}_r)$ the matrix whose columns are the orthonormal eigenvectors of 
$\hat{\boldsymbol{\Gamma}}$ associated to the $r$ largest eigenvalues, $\hat{\lambda}_1 \geq \ldots \geq \hat{\lambda}_r \geq 0$.  Note that these eigenvectors are unique up to sign change and, for $j=1, \ldots, r$, we choose $\hat{\mathbf{v}}_j$ such that $\langle \hat{\mathbf{v}}_j, \mathbf{v}_j \rangle \geq 0$.
Since $\mathbf{v}_1, \ldots, \mathbf{v}_r$ are orthonormal vectors, the spectral norm $\left\| \mathbf{G}_r \right\| =1$. This is also true for $\hat{\mathbf{G}}_r$, and we have $\left\| \hat{\mathbf{G}}_r \right\| =1$.

Now, using the fact that $N^{-1}\hat N=1 + O_p( n^{-1/2})$ and $\| \widehat{\overline{\mathbf{X}}} \| = O_p(p / \sqrt{n})$, it can be shown that 
\begin{align}
\left\| \widehat{\boldsymbol{\Gamma}} - \frac{1}{N} \mathbf{X}^T \mathbf{X} \right\| &= O_p \left( \frac{p}{\sqrt{n}} \right).
\label{err:gammaest}
\end{align}
%We  first show that $N^{-1}\left(\sum_{k\in s}d_k\hat{\mathbf{z}}_{kr}\hat{\mathbf{z}}_{kr}^T-\sum_{k\in U}\mathbf{z}_{kr}\mathbf{z}^T_{kr}\right) =O_p(n^{-1/2})$.
%
%Let $\mathbf{G}=(\mathbf{v}_1, \ldots, \mathbf{v}_p)$ be the $N\times p$ matrix of eigenvectors estimated by $\hat{\mathbf{G}}=(\hat{\mathbf{v}}_1, \ldots, \hat{\mathbf{v}}_p)$. Remark that $\hat{\mathbf{v}}_j$ for $j=1,\ldots, r$ form an orthonormal system, namely $<\hat{\mathbf{v}}_i, \hat{\mathbf{v}}_j>=0$ for $i\neq j$ and $||\hat{\mathbf{v}}_j||=1$. 
We deduce with Lemma 4.3 in Bosq (2000) and equation (\ref{err:gammaest}) that
\begin{align}
\max_{j =1, \ldots, p} | \lambda_j - \hat \lambda_j | & \leq \left\| \widehat{\boldsymbol{\Gamma}} - \frac{1}{N} \mathbf{X}^T \mathbf{X}  \right\| \nonumber \\
 & = O_p \left( \frac{p}{\sqrt{n}} \right),
\label{diff:eigenval}
\end{align}
and
\begin{align}
\left\| \mathbf{v}_j - \hat{\mathbf{v}}_j \right\| &\leq \delta_j \left\| \widehat{\boldsymbol{\Gamma}} - \frac{1}{N} \mathbf{X}^T \mathbf{X} \right\|,
\label{diff:eigenvect}
\end{align}
with $\delta_1 = 2 \sqrt{2}/(\lambda_1 - \lambda_2)$ and $\delta_j = 2 \sqrt{2}/(\min(\lambda_{j-1} - \lambda_j, \lambda_{j} - \lambda_{j+1}))$ for $j=2, \cdots, r$.

Consequently,
\begin{align*}
\left\| \mathbf{G}_r - \hat{\mathbf{G}}_r \right\|^2  &\leq \mbox{tr} \left[ \left(\mathbf{G}_r - \hat{\mathbf{G}}_r \right)^T \left(\mathbf{G}_r - \hat{\mathbf{G}}_r \right) \right] \\
 & \leq \sum_{j=1}^r  \left\| \mathbf{v}_j - \hat{\mathbf{v}}_j \right\|^2 \\
& \leq \sum_{j=1}^r  \delta_j^2 \left\| \widehat{\boldsymbol{\Gamma}} - \frac{1}{N} \mathbf{X}^T \mathbf{X} \right\|^2 \\
 & = O_p \left( \frac{p^2 r^3}{n} \right),
\end{align*}
with (\ref{err:gammaest}), (\ref{diff:eigenvect}) and the fact that $\max_{j=1, \cdots, r} \delta_j^2 = O(r^2)$ which comes from the fact that  we have supposed that $\min_{j=1, \ldots, r+1}( \lambda_{j}-\lambda_{j+1} ) \geq c_\lambda r$ with $c_\lambda >0$.  

We also deduce from (\ref{diff:eigenval}) that 
\begin{align}
\left\| \frac{1}{N} \sum_{k \in U_N} \mathbf{z}_{kr}\mathbf{z}_{kr}^T - \frac{1}{N} \sum_{k \in s} d_k \hat{\mathbf{z}}_{kr} \hat{\mathbf{z}}_{kr}^T  \right\| & \leq \max_{j=1, \ldots, r} | \lambda_j - \hat{\lambda}_j | \nonumber \\
 & = O_p \left( \frac{p}{\sqrt{n}} \right).  
%\left\| \sum_{j=1}^r \lambda_j \mathbf{v}_j \mathbf{v}_j^T - \sum_{j=1}^r \hat{\lambda}_j \hat{\mathbf{v}}_j \hat{\mathbf{v}}_j^T  \right\| \nonumber \\
% & \leq \sum_{j=1}^r | \lambda_j - \hat{\lambda}_j | + 2 \sum_{j=1}^r \lambda_j \left\| \mathbf{v}_j - \hat{\mathbf{v}}_j \right\| \nonumber \\
% & \leq \left(  r + 2 \lambda_r \sum_{j=1}^r \delta_j \right)  \left\| \widehat{\boldsymbol{\Gamma}} - \frac{1}{N} \mathbf{X}^T \mathbf{X} \right\|  \nonumber \\
% &= O_p \left( \frac{p r^2}{\sqrt{n}} \right),
\label{err:gammaestr}
\end{align}
%so that $\sum_{j=1}^r \delta_j = O(r^2).$

Employing a similar technique as before (see the bound obtained in (\ref{maj:difflambda})), we also get  that 
\begin{align}
\left\| \left(\frac{1}{N} \sum_{k \in U_N} \mathbf{z}_{kr}\mathbf{z}_{kr}^T\right)^{-1} - \left(\frac{1}{N} \sum_{k \in s} d_k \hat{\mathbf{z}}_{kr} \hat{\mathbf{z}}_{kr}^T\right)^{-1}  \right\| &= O_p \left( \frac{p}{\sqrt{n}} \right),
\label{err:invgammaest}
\end{align}
and 
\begin{align*}
\left\| \left(\frac{1}{N} \sum_{k \in s} d_k \hat{\mathbf{z}}_{kr} \hat{\mathbf{z}}_{kr}^T\right)^{-1} \right\|^2 &= \frac{1}{\lambda_r^2} + O_p \left( \frac{p^2}{n} \right).
\end{align*}

Using the inequality $\mbox{tr}(\mathbf{AB}) \leq \|\mathbf{A} \| \ \mbox{tr}(\mathbf{B})$ for any symmetric non negative matrices $\mathbf{A}$ and $\mathbf{B}$, we also have 
\begin{align*}
\left\| \frac{1}{N} \sum_{k \in s} d_k \hat{\mathbf{z}}_{kr} y_k  \right\|^2 &= \frac{1}{N^2} \sum_{k,l \in s} d_k d_l y_k y_l \mathbf{x}_l^T \hat{\mathbf{G}}_r^T \hat{\mathbf{G}}_r  \mathbf{x}_k \\
 & = \frac{1}{N^2} \sum_{k,l \in s} d_k d_l y_k y_l \mbox{ tr} \left( \hat{\mathbf{G}}_r^T \hat{\mathbf{G}}_r  \mathbf{x}_l \mathbf{x}_k^T\right) \\
 & = \frac{1}{N^2} \mbox{tr} \left[ \hat{\mathbf{G}}_r^T \hat{\mathbf{G}}_r   \left( \sum_{k,l \in s} d_k d_l y_k y_l \mathbf{x}_l \mathbf{x}_k^T \right)\right] \\
 & \leq \left\| \hat{\mathbf{G}}_r^T \hat{\mathbf{G}}_r \right\| \frac{1}{N^2} \mbox{tr} \left[  \sum_{k,l \in s} d_k d_l y_k y_l \mathbf{x}_l \mathbf{x}_k^T \right] \\
 & \leq \lambda_1 \frac{1}{N} \sum_{k \in s} \left(d_k y_k\right)^2
\end{align*}
because $\left\| \hat{\mathbf{G}}_r^T \hat{\mathbf{G}}_r \right\| = 1$ and the largest eigenvalue of the non negative $N \times N$ matrix ${N}^{-1} \mathbf{X} \mathbf{X}^T$ is equal to $\lambda_1$. Note also that ${N}^{-1} \sum_{k \in s} \left(d_k y_k\right)^2 \leq C$, for some constant $C$,  because  ${N}^{-1} \sum_{k \in U_N} y_k^2$ is supposed to be bounded and  $\max d_k^2 \leq \delta^{-2}$ (see Assumption A2).

We can now bound $\| \hat{\boldsymbol{\gamma}}_{\hat{\mathbf{z}}}(r)  \|$. Combining previous inequalities, we have
\begin{align}
\left\| \hat{\boldsymbol{\gamma}}_{\hat{\mathbf{z}}}(r)  \right\|^2 & \leq \left\| \left(\frac{1}{N} \sum_{k \in s} d_k \hat{\mathbf{z}}_{kr} \hat{\mathbf{z}}_{kr}^T\right)^{-1} \right\|^2 \left\| \frac{1}{N} \sum_{k \in s} d_k \hat{\mathbf{z}}_{kr} y_k  \right\|^2 \nonumber \\
 &= O_p \left(1\right). % + O_p \left( \frac{p^2}{n} \right) \right) \left\| \frac{1}{N} \sum_{k \in s} d_k \hat{\mathbf{z}}_{kr} y_k  \right\|^2
\end{align}

Let us study now $N^{-1} \left(\sum_{k \in U_N} \mathbf{z}_{kr} y_k  -  \sum_{k \in s} d_k \hat{\mathbf{z}}_{kr} y_k \right)$. Writing $\mathbf{z}_{kr} - \hat{\mathbf{z}}_{kr} = \left( \mathbf{G}_r - \hat{\mathbf{G}}_r \right)^T \mathbf{x}_k$, we have %with a similar argument as for inequality (\ref{equ:gammatildebounded}), 
\begin{align*}
\frac{1}{N^2}\left\| \sum_{k \in U_N} \left( \mathbf{z}_{kr} - \hat{\mathbf{z}}_{kr} \right) y_k \right\|^2 & = \frac{1}{N^2} \sum_{k,\ell \in U_N} \mathbf{x}_k^T \left( \mathbf{G}_r - \hat{\mathbf{G}}_r \right)\left( \mathbf{G}_r - \hat{\mathbf{G}}_r \right)^T\mathbf{x}_\ell y_k y_\ell \nonumber \\
&= \frac{1}{N^2} \mbox{tr} \left[ \left( \mathbf{G}_r - \hat{\mathbf{G}}_r \right)\left( \mathbf{G}_r - \hat{\mathbf{G}}_r \right)^T \left( \sum_{k,\ell \in U_N}  \mathbf{x}_\ell  \mathbf{x}_k^Ty_k y_\ell \right) \right] \nonumber \\
 & \leq \left\| \mathbf{G}_r - \hat{\mathbf{G}}_r \right\|^2 \frac{1}{N^2} \left\| \sum_{k \in U_N} \mathbf{x}_k y_k \right\|^2 \nonumber \\
 & \leq \left\| \mathbf{G}_r - \hat{\mathbf{G}}_r \right\|^2  \lambda_1 \left( \frac{1}{N} \sum_{k \in U_N} y_k^2 \right).
\end{align*}
because the largest eigenvalue of the non negative $N \times N$ matrix ${N}^{-1} \mathbf{XG}_r \left(\mathbf{XG}_r\right)^T$ is equal to $\lambda_1$. Since  ${N}^{-1} \sum_{k \in U_N} y_k^2$ is supposed to be bounded, we obtain
\begin{align}
\left\| \frac{1}{N} \left( \sum_{k \in U_N} \left( \mathbf{z}_{kr} - \hat{\mathbf{z}}_{kr} \right) y_k  \right) \right\| = O_p \left( \frac{pr^{3/2}}{\sqrt{n}} \right).
\label{bound:hatzy1}
\end{align}

On the other hand,  noting that $\hat{\mathbf{z}}_{kr} = \hat{\mathbf{G}}_r^T \mathbf{x}_k$ denoting by $\alpha_k = 1 - \ind_{\{k \in s\}} d_k$, we have that 
%with a similar decomposition as in (\ref{bound:z4}) and assumption (A6), we have that $N^{-1} \sum_{k \in U_N} \left\| \hat{z}_{kr} \right\|^4 \leq C_4 r^2$ and 
\begin{align}
\left\| \frac{1}{N} \sum_{k \in U_N} \hat{\mathbf{z}}_{kr} y_k  - \sum_{k \in s} d_k \hat{\mathbf{z}}_{kr} y_k \right\| & = \left\| \hat{\mathbf{G}}_r^T \left( \frac{1}{N} \sum_{k \in U_N} \alpha_k \mathbf{x}_k y_k\right) \right\| \nonumber \\
 & \leq \left\| \hat{\mathbf{G}}_r^T \right\| \left\| \frac{1}{N} \sum_{k \in U_N} \alpha_k \mathbf{x}_k y_k \right\| \nonumber \\
  &= O_p \left( \sqrt{\frac{ p}{n}} \right).
\label{bound:hatzy2}
\end{align}
Combining (\ref{bound:hatzy1}) and (\ref{bound:hatzy2}), we finally obtain that
\begin{align}
\left\| N^{-1} \left(\sum_{k \in U_N} \mathbf{z}_{kr} y_k  -  \sum_{k \in s} d_k \hat{\mathbf{z}}_{kr} y_k \right) \right\| & = O_p \left( \frac{pr^{3/2}}{\sqrt{n}} \right).
\label{bound:hatzy3}
\end{align}

Hence, using now a decomposition similar to (\ref{bound:gammazest}), we obtain 
\begin{align}
\left\| \hat{\boldsymbol{\gamma}}_{\hat{\mathbf{z}}}(r)-\tilde{\boldsymbol{\gamma}}_{\mathbf{z}}(r) \right\| & \leq  \left\| \boldsymbol{\Lambda}_r^{-1} - \left(\frac{1}{N} \sum_{k \in s} d_k \hat{\mathbf{z}}_{kr} \hat{\mathbf{z}}_{kr}^T\right)^{-1} \right\| \left\| \frac{1}{N}  \sum_{k \in U_N} \mathbf{z}_{kr} y_k \right\|  \nonumber \\
  &+ \left\| \frac{1}{N}  \sum_{k \in U_N} \mathbf{z}_{kr} y_k  - \frac{1}{N}  \sum_{k \in s} d_k \hat{\mathbf{z}}_{kr} y_k \right\| \left\| \left(\frac{1}{N} \sum_{k \in s} d_k \hat{\mathbf{z}}_{kr}\hat{\mathbf{z}}_{kr}^T\right)^{-1} \right\| \nonumber \\
&= O_p \left( \frac{pr^{3/2}}{\sqrt{n}} \right).
\label{bound:gammahatzest}
\end{align}
Combining previous bounds we get
\begin{align*}
\left\| \frac{1}{N} \left(\hat{t}_{\mathbf{x}d} -t_{\mathbf{x}} \right)^T \left( \tilde{\boldsymbol{\beta}}^{\rm{pc}}_{\mathbf{x}}(r) - 
\hat{\boldsymbol{\beta}}^{\rm{epc}}_{\mathbf{x}}(r) \right) \right\| & = O_p \left(\sqrt{\frac{p}{n}} \right) O_p \left( \frac{pr^{3/2}}{\sqrt{n}} \right)
\end{align*}
%and, because $\frac{1}{N}\left(\hat{t}_{\hat{\mathbf{z}}_rd}-t_{\hat{\mathbf{z}}_r}\right) = O_p \left(\sqrt{r/n} \right)$, 
%\begin{align}
%\frac{1}{N}\left(\hat{t}_{\hat{\mathbf{z}}_rd}-t_{\hat{\mathbf{z}}_r}\right)^T \left(\hat{\boldsymbol{\gamma}}_{\hat{\mathbf{z}}}(r)-\tilde{\boldsymbol{\gamma}}_{\mathbf{z}}(r) \right) &= O_p \left( \frac{pr^{5/2}}{n} \right).
%\end{align}
%Finally, we clearly have $\frac{1}{N} \left( \hat{t}_{y,\mathbf{x}}^{\rm{diff}}(r) - t_y \right) = O_p \left(\sqrt{r/n} \right)$ 
and using again decomposition (\ref{dec:tzhatdiff}), we finally get 
\begin{align*}
\frac{1}{N} \left( \hat{t}_{yw}^{\rm{epc}} (r) -  t_y \right) &= \frac{1}{N} \left( \tilde{t}_{y,\mathbf{x}}^{\rm{diff}}(r) - t_y \right) + O_p \left( \frac{p^{3/2} r^{3/2}}{n} \right) . 
\end{align*}
% See also the proof of Proposition~\ref{result_pc_estim} below.
\hfill $\square$

%\centerline{(Received xxx 200?; accepted xxx 200?)}\par
\end{document}